\documentclass[twocolumn]{aastex63}

\usepackage{amsmath}
\usepackage{multirow}
\usepackage{hyperref}

\newcommand{\msun}{\ensuremath{M_\odot}}
\newcommand{\mmin}{\ensuremath{m_\mathrm{min}}}
\newcommand{\mmax}{\ensuremath{m_\mathrm{max}}}
\newcommand{\dd}{\ensuremath{\mathrm{d}}}

\newcommand{\truncated}{\textsc{Truncated}}
\newcommand{\Etruncated}{\textsc{Evolving Truncated}}
\newcommand{\binned}{\textsc{Binned Evolution Truncated}}

\newcommand{\brokenpowerlaw}{\textsc{Broken Power Law}}
\newcommand{\Ebrokenpowerlaw}{\textsc{Evolving Broken Power Law}}
\newcommand{\Abrokenpowerlaw}{\textsc{Alternate Evolving Broken Power Law}}

\newcommand{\result}[1]{\textcolor{black}{#1}}

\graphicspath{{./}{figures/}}

\begin{document}

\title{
When are LIGO/Virgo's Big Black-Hole Mergers? \\
}

\author{Maya Fishbach}
\email{maya.fishbach@northwestern.edu}
\affiliation{Center for Interdisciplinary Exploration and Research in Astrophysics (CIERA) and Department of Physics and Astronomy,
Northwestern University, 1800 Sherman Ave, Evanston, IL 60201, USA}

\author{Zoheyr Doctor}
\affiliation{Institute  for  Fundamental  Science, Department of Physics, University of Oregon, Eugene, OR 97403, USA}

\author{Thomas Callister}
\affiliation{Center for Computational Astrophysics, Flatiron Institute, New York, NY 10010, USA}

\author{Bruce Edelman}
\affiliation{Institute  for  Fundamental  Science, Department of Physics, University of Oregon, Eugene, OR 97403, USA}

\author{Jiani Ye}
\affiliation{Department of Physics and Astronomy, Stony Brook University, Stony Brook, NY 11794, USA}

\author{Reed Essick}
\affiliation{Kavli Institute for Cosmological Physics, The University of Chicago, Chicago, IL, 60637, USA}

\author{Will M. Farr}
\affiliation{Center for Computational Astrophysics, Flatiron Institute, New York, NY 10010, USA}
\affiliation{Department of Physics and Astronomy, Stony Brook University, Stony Brook, NY 11794, USA}

\author{Ben Farr}
\affiliation{Institute  for  Fundamental  Science, Department of Physics, University of Oregon, Eugene, OR 97403, USA}

\author{Daniel E. Holz}
\affiliation{Kavli Institute for Cosmological Physics, The University of Chicago, Chicago, IL, 60637, USA}
\affiliation{Enrico Fermi Institute, The University of Chicago, IL, 60637, USA}
\affiliation{Department of Physics, The University of Chicago, IL, 60637, USA}
\affiliation{Department of Astronomy \& Astrophysics, The University of Chicago, IL 60637, USA}

\begin{abstract}
We study the evolution of the binary black hole (BBH) mass distribution across cosmic time. The second gravitational-wave transient catalog (GWTC-2) from LIGO/Virgo contains BBH events out to redshifts $z\sim1$, with component masses in the range $\sim5$--$80\,M_\odot$. In this catalog, the biggest BBHs, with $m_1\gtrsim45\,M_\odot$, are only found at the highest redshifts, $z\gtrsim0.4$. We ask whether the absence of high-mass observations at low redshift indicates that the mass distribution evolves: the biggest BBHs only merge at high redshift, and cease merging at low redshift. Modeling the BBH primary mass spectrum as a power law with a sharp maximum mass cutoff (\truncated~ model), we find that the cutoff increases with redshift ($>99.9\%$ credibility). An abrupt cutoff in the mass spectrum is expected from (pulsational) pair-instability supernova simulations; however, GWTC-2 is only consistent with a Truncated mass model if the location of the cutoff increases from $45^{+13}_{-5}\,M_\odot$ at $z<0.4$ to $80^{+16}_{-13}\,M_\odot$ at $z>0.4$. Alternatively, if the primary mass spectrum has a break in the power law (\brokenpowerlaw) at ${38^{+15}_{-8}\,M_\odot}$, rather than a sharp cutoff, the data are consistent with a non-evolving mass distribution. In this case, the overall rate of mergers at all masses increases with redshift. Future observations will distinguish between a sharp mass cutoff that evolves with redshift and a non-evolving mass distribution with a gradual taper, such as a \brokenpowerlaw. After $\sim100$ BBH merger observations, a continued absence of high-mass, low-redshift events would provide a clear signature that the mass distribution evolves with redshift.
\end{abstract}


\section{Introduction}
\label{sec:intro}

In their first three observing runs, the Advanced LIGO \citep{2015CQGra..32g4001L} and Virgo \citep{2015CQGra..32b4001A} gravitational-wave (GW) detectors observed binary black hole (BBH) mergers out to redshifts $z \sim 1$ \citep{2019PhRvX...9c1040A, 2020arXiv201014527A}.
Observing BBH systems over a range of redshifts allows us to probe the properties of these mergers across cosmic time and unravel how these merging BBH systems came to be.
Previous studies~\citep{2018ApJ...863L..41F,2019ApJ...882L..24A,2020ApJ...896L..32C,2020arXiv200807014R,2020arXiv201014533T,2020arXiv201208839T} have measured the BBH merger rate as a function of redshift, assuming that other properties of the population, including the mass and spin distributions, are constant throughout cosmic time. 
However, there are reasons to expect that the overall BBH population properties may themselves evolve with redshift:
\begin{itemize}
    \item The initial conditions of zero-age main-sequence stars (e.g., metallicity) evolve over cosmic time, which could affect the resultant masses and spins of black holes (BHs) from stellar evolution~\citep{2000ARA&A..38..613K,2010ApJ...715L.138B, 2011A&A...530A.115B,2012ApJ...749...91F,2015ApJ...806..263D,2019ApJ...883L..24S,2019MNRAS.490.3740N, 2020arXiv200906922K, 2020arXiv200906585F, 2020arXiv201011730V}.
    \item If BBH mergers occur in dynamical environments, the dynamical environments could evolve over cosmic time, (dis)favoring mergers of different masses and spins~\citep{Rodriguez:2018rmd,2019MNRAS.482.4528E,2020ApJ...898..152S,2020arXiv201114541R,2021arXiv210102217W}.
    \item The delay time from the inception of a BBH to its merger may depend on the masses and spins of the component BHs and the orbital eccentricity~\citep{2016MNRAS.462..844K,2018PhRvD..97j3014S,2019MNRAS.487....2M}.
    \item Different formation channels (including a formation channel that permits hierarchical mergers) may contribute more or less to the overall merger rate at different times~\citep{Rodriguez:2018rmd,2019PhRvD.100d3027R,2020ApJ...898..152S,2020ApJ...896..138Y,2020arXiv201110057Z}.
\end{itemize}
The combined effect of these phenomena would manifest in the GW data as mass and/or spin distributions of merging black holes that are different at different redshifts.
For example, Fig.~1 of \citet{Rodriguez2019} shows that the mass distribution for mergers in dense star clusters extends to higher masses when considering mergers at all redshifts compared to mergers with $z<1$.
Similarly, some population synthesis models of BBH mergers from isolated binary evolution exhibit more support for higher mass mergers at higher redshifts, although this effect is expected to be mild at the redshifts accessible to Advanced LIGO and Advanced Virgo~\citep[e.g.][]{2019MNRAS.487....2M}.
With a growing catalog of BBH events, we can begin to empirically measure the existence (or lack thereof) of redshift evolution out to $z\sim1$.

The second gravitational-wave transient catalog \citep[GWTC-2;][]{2020arXiv201014527A} contains 44 confident BBH mergers with $m_1 > m_2 > 3\,\msun$ and a false-alarm rate (FAR) below $1\,\mathrm{yr}^{-1}$.
While the first LIGO/Virgo catalog release, GWTC-1, found evidence for a dearth of systems with $m_1 > 45\,\msun$, the updated GWTC-2 catalog contains several systems with $m_1 > 45\,\msun$.
The presence of these high-mass black holes leads \citet{2020arXiv201014533T} to infer that the primary mass spectrum is more complicated than a power law with an abrupt mass cutoff.
Although a mass cutoff at $\sim 40\,\msun$ was consistent with GWTC-1, \citet{2020arXiv201014533T} concluded that the GWTC-2 observations are more consistent with a \emph{break} or a \emph{peak} at $\sim 40\,\msun$, with the mass distribution declining more steeply at higher masses.

\begin{figure*}
    \centering
    \includegraphics[width=1.0\textwidth]{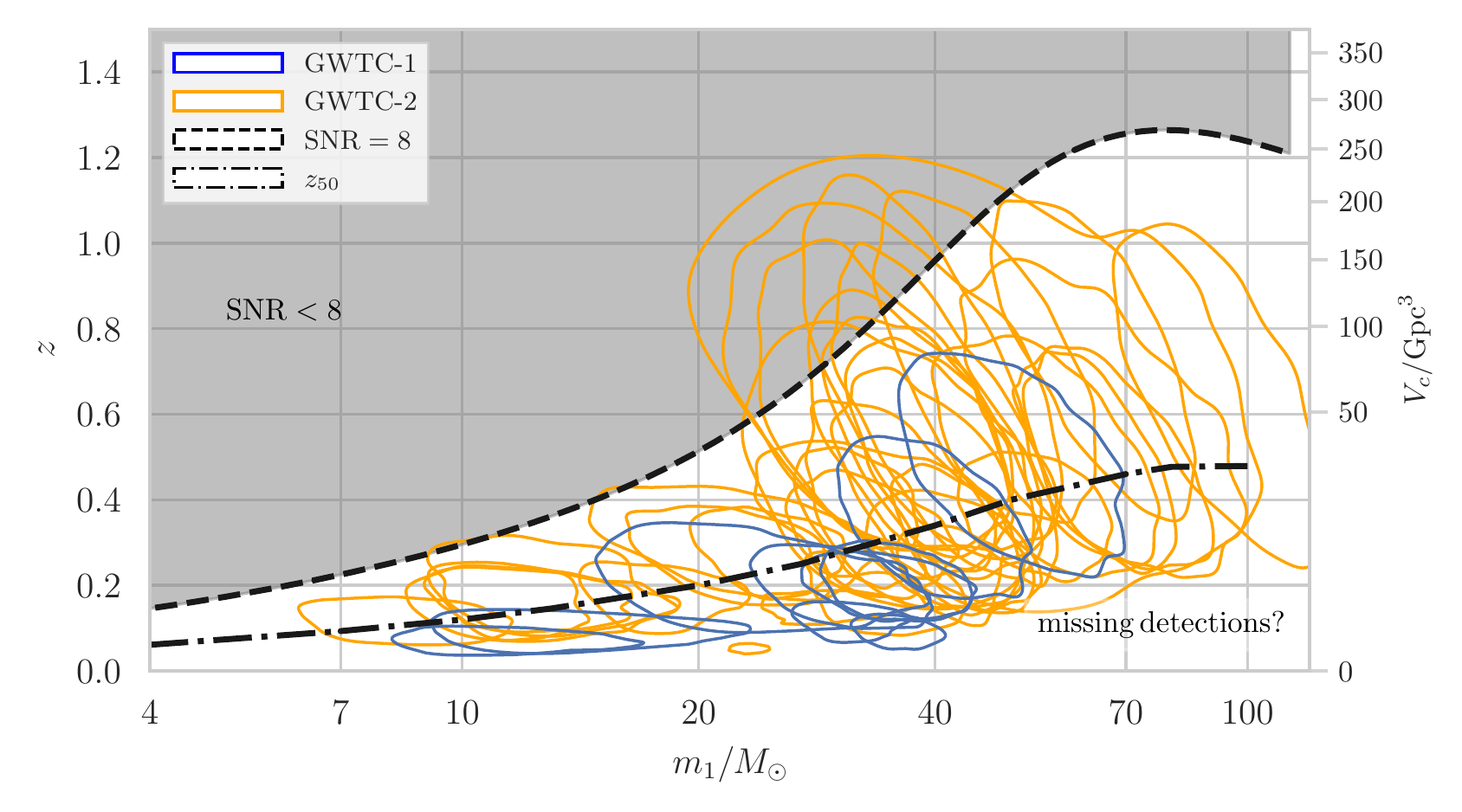}
    \caption{
        Posterior probability distribution of the primary mass $m_1$ and redshift $z$ for the 34 confident BBH events announced in GWTC-2 (\emph{orange}) and 10 confident BBHs in GWTC-1 (\emph{blue}).
        The contours enclose 90\% of the posterior probability, inferred under the default parameter estimation priors used in the GWTC-1 and GWTC-2 publications \citep[][]{2019PhRvX...9c1040A,2020arXiv201014527A}.
        The shaded region denotes the redshifts for which the signal-to-noise ratio (SNR) of an equal-mass BBH is less than 8 for an optimally-oriented component BH mass under the Advanced LIGO ``mid-high" noise curve \citep{Aasi:2013wya}; we do not detect low-mass, high-redshift events because they are too quiet. As a point of reference, we overlay the curve representing $z_\mathrm{50}$, the expected median distance of detected mergers that are distributed with constant rate density per co-moving volume.
        The question of whether there are equal numbers of events above and below this $z_\mathrm{50}$ line motivates us to examine whether the intrinsic BBH mass distribution evolves with redshift.
    }
    \label{fig:m1-z-catalog}
\end{figure*}

Notably, all of the high-mass observations in GWTC-2 are also at relatively high redshifts, as seen in Fig.~\ref{fig:m1-z-catalog}, with a noticeable absence of events with $m_1 \gtrsim 45\,\msun$ at low redshifts. 
For reference, we also overlay the expected median redshift $z_{50}$ of sources distributed with a constant rate per comoving volume\footnote{$z_{50}$ is computed via \url{https://users.rcc.uchicago.edu/~dholz/gwc/}, an online calculator based on \citet{Chen:2017wpg}, assuming equal mass mergers and the ``Advanced LIGO mid-high" noise curve, which can be found here: \url{https://dcc.ligo.org/public/0094/P1200087/019/fig1_aligo_sensitivity.txt}}~\citep{Chen:2017wpg,Aasi:2013wya}.
The apparent absence is not necessarily surprising, because the highest mass systems are also detectable at the highest redshifts, and, if these systems are rare, we expect to detect them primarily at high redshifts where there is more cosmological volume.
However, this also suggests an alternative explanation for the high-mass detections in GWTC-2, also proposed by~\citet{2019ApJ...883L..24S}: the underlying astrophysical mass distribution may skew to higher masses at higher redshifts, implying that a higher fraction of high mass mergers (per comoving time-volume) occur at higher redshifts than at low redshifts.
We explore this hypothesis in more detail below.

The remainder of the paper is structured as follows. Section~\ref{sec:methods} introduces the phenomenological models that we use to describe the BBH population, and the statistical framework that we use to fit the models given the GWTC-2 data. Section~\ref{sec:results} presents our main results regarding the evolution of the BBH mass distribution. We carry out posterior predictive checks in Section~\ref{sec:checks}, and discuss future prospects before concluding in Section~\ref{sec:conclusion}.


\section{Methods}
\label{sec:methods}

We first describe the parametrized population models we assume in Section~\ref{sec:models} before laying out how we use these within the statistical framework of Section~\ref{sec:stat}.

\subsection{Population models}
\label{sec:models}

\begin{table*}
    \begin{center}
    \begin{tabular}{m{1.6cm} ccc}
        \hline
        \textbf{Name} & \textbf{Nonevolving Parameters} & \multicolumn{2}{c}{\textbf{Evolving Parameters}} \\
        \hline\hline
            & $m_\mathrm{min}\sim \mathcal{U}(2\,\msun, 10\,\msun)$ & \multicolumn{2}{c}{\multirow{6}{*}{None}} \\
            & $m_\mathrm{max} = 100 M_\odot$ & & \\
        \multirow{2}{*}{\truncated}
            & $m_\mathrm{break} \sim \mathcal{U}(30\,\msun, 100\,\msun)$ & & \\
            & $\alpha_1 \sim \mathcal{U}(-5, 2)$ & & \\
            & $\alpha_2 = -20$ & & \\
            & $\beta \sim \mathcal{U}(-4, 12)$ & & \\
        \cline{2-4}
            & & \multirow{2}{*}{$m_\mathrm{min}(z) = m_\mathrm{min}^\mathrm{low} + (m_\mathrm{min}^\mathrm{high}-m_\mathrm{low}^\mathrm{high})\Theta(z>z_0)$} & $m_\mathrm{min}^\mathrm{low} \sim \mathcal{U}(2\,\msun, 10\,\msun)$ \\
            & & & $m_\mathrm{min}^\mathrm{high} \sim \mathcal{U}(2\,\msun, 10\,\msun)$ \\
        \cline{3-4}
        \textsc{Binned}
            & $m_\mathrm{max} = 100 M_\odot$ & \multirow{2}{*}{$m_\mathrm{break}(z) = m_\mathrm{break}^\mathrm{low} + (m_\mathrm{break}^\mathrm{high}-m_\mathrm{break}^\mathrm{low})\Theta(z>z_0)$} & $m_\mathrm{break}^\mathrm{low} \sim\mathcal{U}(30\,\msun, 100\,\msun)$ \\
        \textsc{Evolution}
            & $\alpha_2 = -20$ & & $m_\mathrm{break}^\mathrm{high} \sim\mathcal{U}(30\,\msun, 100\,\msun)$ \\
        \cline{3-4}
        \textsc{Truncated}
            & $z_0 = 0.4$ & \multirow{2}{*}{$\alpha_1(z) = \alpha_1^\mathrm{low} + (\alpha_1^\mathrm{high}-\alpha_1^\mathrm{low})\Theta(z>z_0)$} & $\alpha_1^\mathrm{low} \sim \mathcal{U}(-5, 2)$ \\
            & & & $\alpha_1^\mathrm{high} \sim \mathcal{U}(-5, 2)$ \\
        \cline{3-4}
            & & \multirow{2}{*}{$\beta_z(z) = \beta^\mathrm{low} + (\beta^\mathrm{high}-\beta^\mathrm{low})\Theta(z>z_0)$} & $\beta^\mathrm{low} \sim \mathcal{U}(-4, 12)$ \\
            & & & $\beta^\mathrm{high} \sim \mathcal{U}(-4, 12)$ \\
        \cline{2-4}
            & $m_\mathrm{min} \sim \mathcal{U}(2\,\msun, 10\,\msun)$ & & \\
        \multirow{2}{*}{\textsc{Evolving}}
            & $m_\mathrm{max} = 100 M_\odot$ & \multirow{2}{*}{$m_\mathrm{break}(z) = m_\mathrm{min} + (m_\mathrm{max}-m_\mathrm{min}) b(z)$} & \multirow{2}{*}{$b_0 \sim \mathcal{U}(0, 1)$} \\
        \multirow{2}{*}{\textsc{Truncated}}
            & $\alpha_1 \sim \mathcal{U}(-5, 2)$ & \multirow{2}{*}{$b(z) = (1 + (b_0^{-1}-1)\exp(-b_1 z))^{-1}$} & \multirow{2}{*}{$b_1 \sim \mathcal{U}(-4, 4)$} \\
            & $\alpha_2 = -20$ & & \\
            & $\beta \sim \mathcal{U}(-4, 12)$ & & \\
        \hline
            & $m_\mathrm{min}\sim \mathcal{U}(2\,\msun, 10\,\msun)$ & \multicolumn{2}{c}{\multirow{6}{*}{None}} \\
        \multirow{2}{*}{\textsc{Broken}}
            & $m_\mathrm{max} \sim \mathcal{U}(65\,\msun, 100\,\msun)$ & & \\
        \multirow{2}{*}{\textsc{Power}}
            & $m_\mathrm{break} \sim \mathcal{U}(20\,\msun, 65\,\msun)$ & & \\
        \multirow{2}{*}{\textsc{Law}}
            & $\alpha_1 \sim \mathcal{U}(-5, 2)$ & & \\
            & $\alpha_2 \sim \mathcal{U}(-12, 2)$ & & \\
            & $\beta \sim \mathcal{U}(-4, 12)$ & & \\
        \cline{2-4}
        \multirow{2}{*}{\textsc{Evolving}}
            & $m_\mathrm{min}\sim \mathcal{U}(2\,\msun, 10\,\msun)$ & & \\
        \multirow{2}{*}{\textsc{Broken}}
            & $m_\mathrm{max} \sim \mathcal{U}(65\,\msun, 100\,\msun)$ & \multirow{2}{*}{$m_\mathrm{break}(z) = m_\mathrm{min} + (m_\mathrm{max}-m_\mathrm{min}) b(z)$} & \multirow{2}{*}{$b_0 \sim \mathcal{U}(0, 1)$} \\
        \multirow{2}{*}{\textsc{Power}}
            & $\alpha_1 \sim \mathcal{U}(-5, 2)$ & \multirow{2}{*}{$b(z) = (1 + (b_0^{-1}-1)\exp(-b_1 z))^{-1}$} & \multirow{2}{*}{$b_1 \sim \mathcal{U}(-4, 4)$} \\
        \multirow{2}{*}{\textsc{Law}}
            & $\alpha_2 \sim \mathcal{U}(-10, 2)$ & & \\
            & $\beta \sim \mathcal{U}(-4, 12)$ & & \\
        \cline{2-4}
        \multirow{2}{*}{\textsc{Alternate}}
            & $m_\mathrm{min}\sim \mathcal{U}(2\,\msun, 10\,\msun)$ & & \\
        \multirow{2}{*}{\textsc{Evolving}}
            & $m_\mathrm{max} \sim \mathcal{U}(65\,\msun, 100\,\msun)$ & \multirow{3}{*}{$\alpha_2(z) = \alpha_2^0 + \alpha_2^\prime z$} & \multirow{2}{*}{$\alpha_2^0 \sim \mathcal{U}(-10, 2)$} \\
        \multirow{2}{*}{\textsc{Broken}}
            & $m_\mathrm{break} \sim \mathcal{U}(20\,\msun, 65\,\msun)$ & & \multirow{2}{*}{$\alpha_2^\prime \sim \mathcal{U}(-12, -12)$} \\
        \multirow{2}{*}{\textsc{Power Law}}
            & $\alpha_1 \sim \mathcal{U}(-5, 2)$ & & \\
            & $\beta \sim \mathcal{U}(-4, 12)$ & & \\
        \hline
    \end{tabular}
    \end{center}
    \caption{
        Prior ranges for mass models conditioned on redshift: $p(m_1, m_2|z) = p(m_2|m_1,z) p(m_1|z)$ described by Eqs~\ref{eq:prob(m1)} and~\ref{eq:prob(q)}.
        For each model considered, we separate the parameters that do not evolve with redshift from those that do, showing the assumed functional forms of the evolution.
        $\Theta(\cdot)$ represents the Heaviside function.
        All hyperparameters are drawn from uniform priors between $\mathcal{X}$ and $\mathcal{Y}$: $\mathcal{U}(\mathcal{X}, \mathcal{Y})$.
    }
    \label{tab:models overview}
\end{table*}

\begin{table}
    \begin{center}
    \begin{tabular}{ll}
        \hline
         $f(z) = (1+z)^\kappa$        & $\kappa \sim \mathcal{U}(-6, 6)$ \\
         $p(\chi_\mathrm{eff}) \propto \exp\left[-(\chi_\mathrm{eff}-\mu^2/2\sigma^2\right]$
           & $\mu \sim \mathcal{U}(-0.5, 0.5)$ \\
         \quad\quad\quad\quad\quad\quad $\times \Theta(-1 \leq \chi_\mathrm{eff} \leq 1)$ & $\sigma \sim \mathcal{U}(0.02, 1)$ \\
         \hline
    \end{tabular}
    \end{center}
    \caption{
        Prior ranges assumed for $f(z)$ and $p(\chi_\mathrm{eff})$.
        Together with one of the models specified in Table~\ref{tab:models overview}, these distributions form our population model for the instantaneous rate-density in the source-frame $d\mathcal{R}/dm_1 dm_2 d\chi_\mathrm{eff}$ (Eq.~\ref{eq:diff-rate-general}).
        $\Theta(\cdot)$ represents the Heaviside function.
        All hyperparameters are drawn from uniform priors between $\mathcal{X}$ and $\mathcal{Y}$: $\mathcal{U}(\mathcal{X}, \mathcal{Y})$.
    }
    \label{tab:f(z) and p(spin)}
\end{table}

We use simple phenomenological parametric models to describe the distribution of BBH masses $m_1$ and $m_2$, effective spin $\chi_\mathrm{eff}$, and redshift $z$, based on the models used in~\citet{2020arXiv201014533T}.
We write the differential merger rate density (number of mergers per comoving volume per source-frame time) as:
\begin{equation} \label{eq:diff-rate-general}
    \frac{\dd\mathcal{R}(m_1, m_2, \chi_\mathrm{eff}, z)}{\dd m_1 \dd m_2 \dd \chi_\mathrm{eff}} = \mathcal{R}_0 p(m_1, m_2 \mid z) p(\chi_\mathrm{eff}) f(z),
\end{equation}
where $\mathcal{R}_0$ is the rate density at redshift $z = 0$, $p(m_1, m_2 \mid z)$ is the two-dimensional source-frame mass distribution at a given redshift, $p(\chi_\mathrm{eff})$ is the distribution of effective spins (assumed to be independent of $z$), and $f(z)$ describes the evolution of the overall merger rate with redshift.
The probability distributions $p$ are normalized, so that $p(m_1, m_2 \mid z)$ integrates to unity over the allowed $m_1$, $m_2$ (here taken to be $2\,\msun < m_2 < m_1 < 100\,\msun$ to match the mass range of the simulated detections used to estimate searches' sensitivities), and $p(\chi_\mathrm{eff})$ integrates to unity over $-1 < \chi_\mathrm{eff} < 1$.
The function $f(z)$ is chosen so that $f(z = 0) = 1$.
Thus, integrating the differential merger rate $\dd\mathcal{R} / \dd m_1 \dd m_2 \dd \chi_\mathrm{eff}$ over all masses and spins at a given $z$ gives the overall merger rate density at that redshift.
The rate density of Eq.~\ref{eq:diff-rate-general} can alternatively be written in terms of the number density, defined by the relation:
\begin{equation}
    \frac{\dd \mathcal{R}}{\dd m_1 \dd m_2 \dd \chi_\mathrm{eff}} = \frac{\dd N}{\dd m_1 \dd m_2 \dd \chi_\mathrm{eff} \dd V_c \dd t_\mathrm{src}},
\end{equation}
where $V_c$ is the comoving volume element~\citep{1999astro.ph..5116H} and $t_\mathrm{src}$ is the time measured in the source frame, so that after integrating over the observing time in the detector frame:
\begin{equation}
    \frac{\dd N (m_1, m_2, \chi_\mathrm{eff}, z)}{\dd m_1 \dd m_2 \dd \chi_\mathrm{eff} \dd z} = \frac{\dd V_c}{\dd z} \left(\frac{T_\mathrm{obs}}{1 + z}\right) \frac{\dd \mathcal{R}}{\dd m_1 \dd m_2 \dd \chi_\mathrm{eff}},
\end{equation}
where $T_\mathrm{obs}$ is the total observing time and the factor of $(1 + z)$ converts source-frame time to detector-frame time. Integrating the number density ${\dd N} / {\dd m_1 \dd m_2 \dd \chi_\mathrm{eff} \dd z}$ over all masses, spins and redshifts up to some maximum $z_\mathrm{max}$ gives the total expected number of BBH mergers in the universe out to $z_\mathrm{max}$.

With the factorization of Eq.~\ref{eq:diff-rate-general}, our model for the joint mass-spin-redshift distribution of BBH systems consists of a redshift-dependent mass distribution, a spin distribution, and a rate evolution function.
We introduce a parametric model and a set of hyperparameters to describe each of these components, listed in Tables~\ref{tab:models overview} and \ref{tab:f(z) and p(spin)}.
Exploring the redshift-dependence of the mass distribution is the focus of this work, so we consider a few different parametric forms for the mass distribution $p(m_1, m_2 \mid z)$, but fix the parametric form of the spin distribution and rate evolution.
For the rate evolution, we assume $f(z) = (1 + z)^\kappa$ following \citet{2018ApJ...863L..41F, 2019ApJ...882L..24A, 2020arXiv201014533T}.
For the spin distribution, we assume a truncated Gaussian distribution for the effective spin during the inspiral $\chi_\mathrm{eff}$, described by a mean $\mu_\mathrm{eff}$ and standard deviation $\sigma_\mathrm{eff}$, and truncated to the physical range $[-1, 1]$ \citep{2020ApJ...895..128M, 2019MNRAS.484.4216R, 2020arXiv201014533T}.
We ignore other spin degrees of freedom because only $\chi_\mathrm{eff}$ correlates noticeably with our mass and redshift inference~\citep{2018PhRvD..98h3007N}.

We use two different underlying parametric distributions for the component masses.
\brokenpowerlaw, similar to the model defined in \citet{2020arXiv201014533T}, models the primary mass distribution as a power law from $m_\mathrm{min}$ to $m_\mathrm{max}$ with an additional parameter $m_\mathrm{break}$ where the power law spectral index changes from $\alpha_1$ to $\alpha_2$:
\begin{equation}\label{eq:prob(m1)}
    p(m_1) \propto \left\{ \begin{matrix}
    \left(m_1 / m_\mathrm{break}\right)^{\alpha_1} & \mathrm{if} \ m_\mathrm{min} \leq m_1 < m_\mathrm{break} \\ \left(m_1/m_\mathrm{break}\right)^{\alpha_2} & \mathrm{if} \ m_\mathrm{break} \leq m_1 < m_\mathrm{max} \\
    0 & \mathrm{else}
    \end{matrix} \right.
\end{equation}
We also modify this model to approximate the \truncated~model also defined in \citet{2020arXiv201014533T} by fixing the second power law exponent $\alpha_2 = -20$, which mimics the hard cutoff of the \truncated~model.
In our approximated \truncated~model, $m_\mathrm{break}$ becomes the high-mass cutoff, and we fix $\mmax = 100\,\msun$.
Eq.~\ref{eq:prob(m1)} describes both these models, and specific prior ranges as well as assumed functional forms of the distributions' evolution with redshift are given in Table~\ref{tab:models overview}.
For simplicity, we adopt a sharp lower bound on our mass distributions instead of the tapering function employed in~\citet{2020arXiv201014533T}, as this has no effect on the high-mass inference on which we focus in this work.

In both of these models we describe the conditional mass ratio distribution with a single power law with index $\beta$:
\begin{multline}\label{eq:prob(q)}
    p(q|m_1) = m_1 p(m_2|m_1)
        = \left(\frac{\beta+1}{1 - \left(\mmin/m_1\right)^{\beta+1}}\right) q^\beta
\end{multline}
Assumed prior ranges for $\beta$ are also shown in Table~\ref{tab:models overview}.


\subsection{Statistical framework}
\label{sec:stat}
We use hierarchical Bayesian inference to fit these models, marginalizing over individual event properties and the expected number of detections during the observing period~\citep{2004AIPC..735..195L,2010PhRvD..81h4029M,2019MNRAS.486.1086M}.
Given data $\{d_i\}$ from $N_\mathrm{det}$ GW events, we wish to infer the parameters describing our chosen population distributions $\Lambda$.
Using Bayes' rule, we can  write out the full hierarchical posterior distribution as:
\begin{widetext}
\begin{equation} \label{eq:Posterior}
    p\left(\Lambda, \mathcal{R}_0 | \{d_i\} \right) \propto \mathcal{R}_0^{N_\mathrm{det}} e^{-\mathcal{R}_0\xi(\Lambda)} \Bigg[\prod_{i=1}^{N_\mathrm{det}} \int \mathcal{L}(d_i|m_1,m_2,\chi_\mathrm{eff},z) p \left(m_1, m_2, \chi_\mathrm{eff}, z | \Lambda \right) dm_1 dm_2 d\chi_\mathrm{eff} dz \Bigg] p( \Lambda, \mathcal{R}_0)
\end{equation}
Here we define the $i^\mathrm{th}$ individual event likelihood as $\mathcal{L}(d_i|m_1,m_2,\chi_\mathrm{eff},z)$ and $\xi(\Lambda)$ as the fraction of binary sources we would expect to successfully detect for a given population model defined by the hyperparameters $\Lambda$:
\begin{equation}
\label{eq:xi}
    \xi(\Lambda) = \int p(m_1, m_2, \chi_\mathrm{eff}, z | \Lambda) P(\mathrm{det}|m_1, m_2, \chi_\mathrm{eff}, z) \dd m_1 \dd m_2 \dd \chi_\mathrm{eff} \dd z
\end{equation}
where $P(\mathrm{det}|m_1, m_2, \chi_\mathrm{eff}, z)$ is the probability of detecting a single system with parameters $m_1$, $m_2$, $\chi_\mathrm{eff}$, and $z$.
Assuming a log-uniform prior on $\mathcal{R}_0$, we marginalize and write the posterior as:
\begin{align}
    p\left(\Lambda\right | \{d_i\}) & \propto \frac{p(\Lambda)}{\xi(\Lambda)^{N_\mathrm{det}}} \prod_{i=1}^{N_\mathrm{det}} \int \mathcal{L}(d_i|m_1,m_2,\chi_\mathrm{eff},z) p\left(m_1, m_2, \chi_\mathrm{eff}, z | \Lambda \right) dm_1 dm_2 d\chi_\mathrm{eff} dz \label{eq:MargPosterior} \\
    & \approx \frac{p(\Lambda)}{\xi(\Lambda)^{N_\mathrm{det}}} \prod_{i=1}^{N_\mathrm{det}} \left(\frac{1}{K_i} \sum_{j}^{K_i} \frac{p(m_1^{ij}, m_2^{ij}, \chi_\mathrm{eff}^{ij}, z^{ij} | \Lambda )}{\pi(m_1^{ij},m_2^{ij},\chi_\mathrm{eff}^{ij},z^{ij})} \right) \label{eq:AvgMargPosterior}
\end{align}
\end{widetext}
where we approximate the integrals over single-event parameters via importance sampling with $K_i$ single-event posterior samples generated with prior $\pi(m_1,m_2,\chi_\mathrm{eff},z)$: $\{m_1^{ij}, m_2^{ij}, \chi_\mathrm{eff}^{ij}, z^{ij}\}$ denoting the $j^\mathrm{th}$ sample for the $i^\mathrm{th}$ event. The single-event posterior samples for the 44 events used in this analysis are taken from~\citet{2019arXiv191211716T} and~\citet{O3a_pe}. To compute the sum in Eq.~\ref{eq:AvgMargPosterior}, we use the same set of samples, derived under the same priors and waveform models, as in~\citet{2020arXiv201014533T}.

We additionally approximate $\xi(\Lambda)$ via importance sampling over sets of detected simulated events,\footnote{The simulated detection sets covering the O3a observing run can be found at \url{https://dcc.ligo.org/LIGO-P2000217/public}. For the first two observing runs, we use the mock injection sets used in~\citet{2020arXiv201014533T}, which can be found at \url{https://dcc.ligo.org/LIGO-P2000434/public}.} marginalizing over our uncertainty from the finite number of simulated events available~\citep{Farr2019}.

Given this population likelihood, we sample from the posterior on the population hyperparameters $\Lambda$ using the Monte-Caro samplers \textsc{PyMC3}~\citep{2016ascl.soft10016S}, \textsc{emcee}~\citep{2013ascl.soft03002F}, and 
\textsc{Stan}~\citep{2017JSS....76....1C}

Once we have posterior samples $\Lambda$, we perform a series of posterior predictive checks.
In essence, the posterior predictive checks are goodness-of-fit checks that compare the observed data and synthetic events drawn from the fits to the population models.
If the models correctly account for all variation within the observations, synthetic ``predicted events'' drawn from the population hyperposterior should resemble the observed behavior of real events.
The first example of such a check (Fig.~\ref{fig:m1_versus_z_truncated}) is introduced in the following Section~\ref{sec:truncatedPPC}.

\section{Constraints on the evolution of the mass distribution}
\label{sec:results}

We describe the consistency between the data and a series of increasingly complex population models in what follows, beginning with a simple non-evolving \truncated~power law in Section~\ref{sec:truncatedPPC} and then proceeding to more complicated models that do evolve with redshift in Sections~\ref{sec:binned} and~\ref{sec:smooth-evol}.
We also confirm the robustness of our conclusions to a few possible systematics in Sections~\ref{sec:GW190521} and~\ref{sec:selection}.


\subsection{Tension with the non-evolving \truncated\ model}
\label{sec:truncatedPPC}

\begin{figure}
    \centering
    \includegraphics[width = 0.49\textwidth]{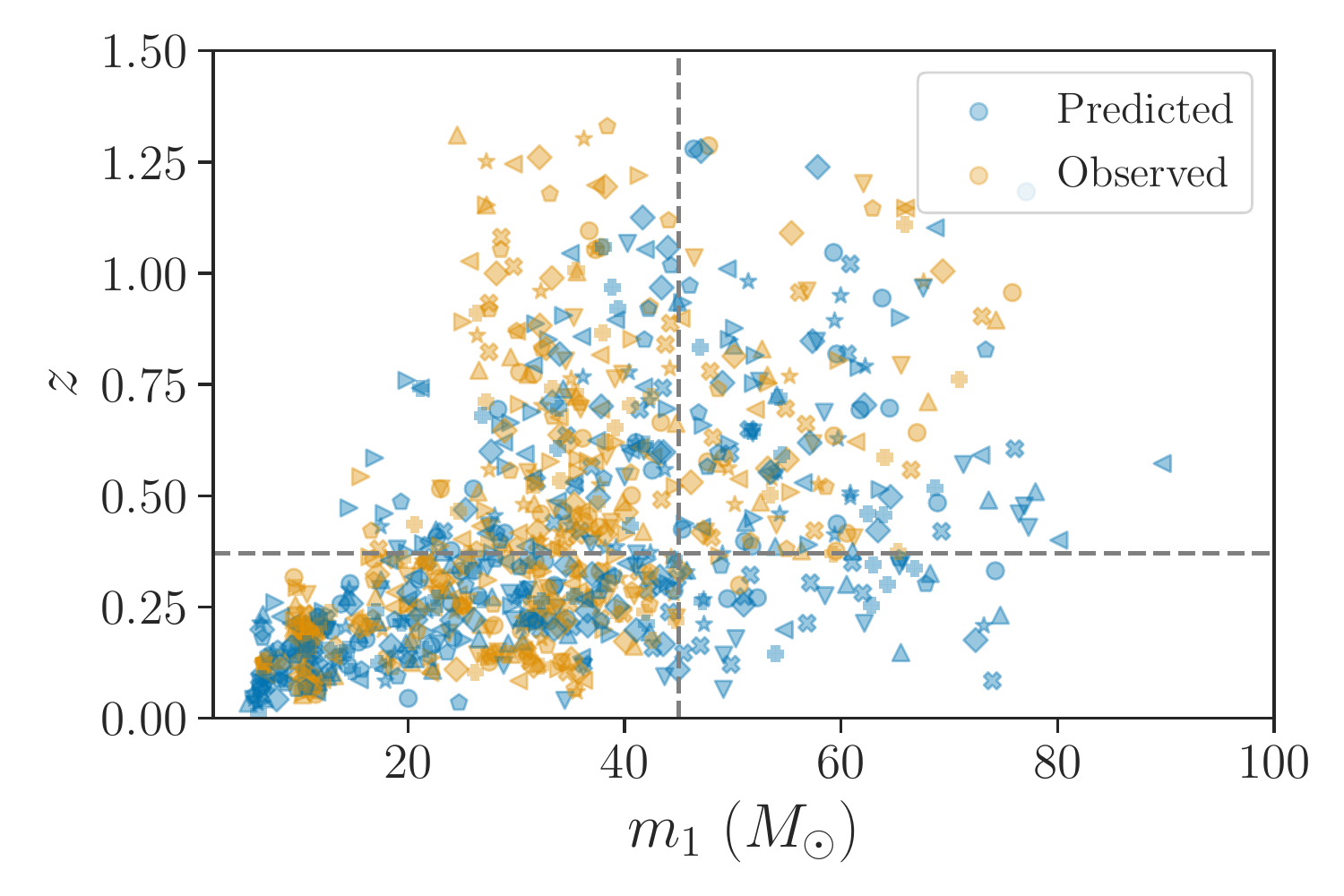}
    \caption{
        Primary masses and redshifts of the 44 confident BBH observations (\emph{orange}) compared to 44 draws from the predicted observable distribution (\emph{blue}), inferred under the non-evolving \truncated~model.
        Each marker shape corresponds to a different set of 44 draws, where each set is inferred under a different population model drawn from the hyperposterior.
        We plot 10 total sets.
        The dashed lines denote $z = 0.37$ (the median observed redshift in the sample) and $m_1 = 45\,\msun$.
        The top left corner (low masses, high redshifts) contains no predicted or observed events because it is beyond the detection horizon. On the other hand, the bottom right corner (high masses, low redshifts) contains predicted events, but no observed events, showing that the model generally overpredicts the number of high-mass ($m_1 > 45\,\msun$) events at low redshifts ($z < 0.37$) compared to our observations.
    }
    \label{fig:m1_versus_z_truncated}
\end{figure}

To see if the dearth of high-mass events at low redshifts shown in Fig.~\ref{fig:m1-z-catalog} is statistically significant, we first present a posterior predictive check using the non-evolving \truncated\ model.  We use the statistical methods described above to fit the \truncated\ model (see Table~\ref{tab:models overview}) to the GWTC-2 events. 
Fig.~\ref{fig:m1_versus_z_truncated} shows the observed primary masses and redshifts of BBH events (orange), compared to the prediction from the non-evolving \truncated~model (blue).
For every hyperposterior sample in the \truncated~model, we draw 44 synthetic events from the predicted distribution and record their primary masses and redshifts.
We likewise draw one primary mass and redshift sample for each of the 44 events in our catalog, inferred under the same draw from the population hyperposterior.
Fig.~\ref{fig:m1_versus_z_truncated} shows this comparison for 10 sets of fair draws from the population hyperposterior, each set marked with  a distinct symbol.
We note that the GWTC-2 events with $m_1 \gtrsim 45\,\msun$ occur at redshifts $z \gtrsim 0.37$.
The model tends to overpredict the maximum observed mass at low redshift in order to match the maximum observed mass at high redshift.
Splitting each set of predicted and observed events at their median redshifts to define ``low" and ``high" redshift events, the \truncated\ model overpredicts the largest mass seen at low redshifts \result{91\%} of the time,
typically overestimating the observed maximum mass at low redshifts by \result{$18^{+18}_{-22}\,\msun$}. 
On the other hand, at high redshifts, the predicted maximum mass typically matches the observed maximum mass for both models, with an average difference of only $\sim 2\,\msun$ between the predicted and observed maximum mass for the \truncated\ model.

This baseline analysis shows that the mismatch between predicted and observed masses and redshifts suggested by Fig.~\ref{fig:m1-z-catalog} also appears when we consider an overly simple population distribution. A similar conclusion regarding the failure of the \truncated~model to fit the GWTC-2 data was found by \citet{2020arXiv201014533T}, who pointed out the tension between the observed primary mass distribution and the \truncated~model prediction. Fig.~\ref{fig:m1_versus_z_truncated} recasts this tension in terms of the joint mass and redshift distribution, and corroborates our expectation from Fig.~\ref{fig:m1-z-catalog} that a redshift-dependent mass distribution may be needed to accurately describe the observed population.

\subsection{Two redshifts bins}
\label{sec:binned}

To further explore whether the data support a different mass distribution at high redshifts compared to low redshifts, we first perform a change-point analysis.
We fit the mass distribution in two redshift bins ($z < 0.4$ and $z > 0.4$), which splits the events roughly evenly between the two bins.
We assume that the \truncated~model describes the mass distribution in both bins, but we allow the parameters describing the mass distribution to jump discontinuously between the bins. We refer to this model as the \binned~model (see Table~\ref{tab:models overview} for prior ranges).
Unsurprisingly, we find a strong preference that the maximum mass in the high-redshift bin is larger than the maximum mass in the low-redshift bin (\result{99.4\%} credibility).
This preference remains (\result{95.2\%} credibility) even when we exclude the most massive event in the high-redshift bin: GW190521 (see Section~\ref{sec:GW190521}).
Including GW190521, we infer that the maximum mass in the high-redshift bin is larger by \result{${35}^{+17}_{-17}\,\msun$} than the maximum mass in the low-redshift bin. Without GW190521, the high-redshift maximum mass is larger than the low-redshift maximum mass by \result{$ 16^{+12}_{-15}\,\msun$}.  

Meanwhile, the other parameters describing the mass distribution are consistent between the two bins, albeit with large uncertainties.
For example, the power-law slope of the mass distribution in the high redshift bin is poorly constrained, because it is degenerate with the redshift evolution of the merger rate.
Steep (negative) $m_1$ power-law slopes correspond to steep (positive) redshift evolution slopes, because steeper mass distributions with fewer high-mass events must have a larger overall merger rate at high redshift to support the number of high-mass events observed at large redshifts~\citep{2018ApJ...863L..41F}.
In the following, we focus on the high-mass end of the mass distribution and its possible evolution with redshift.

\subsection{Continuous evolution with redshift}
\label{sec:smooth-evol}

Motivated by the \binned~analysis, we next allow the high-mass end of the mass distribution to evolve continuously and monotonically with redshift.
We parameterize the location of the break in the power law as a function of redshift, allowing it to vary between $2\,\msun < \mmin < m_\mathrm{break} < \mmax < 100\,\msun$.

We consider two scenarios for the evolving mass distribution: an \Etruncated~model in which the high-mass slope is fixed to be steep ($\alpha_2 = -20$) to approximate a sharp cutoff that evolves with redshift, and an \Ebrokenpowerlaw~model in which we fit for the high-mass slope, considering values $\alpha_2 > -10$.
Table~\ref{tab:models overview} shows the functional form of the models and their full prior ranges.
For the \Etruncated~model, $m_\mathrm{break}(z)$ parameterizes the location of the cutoff.

\begin{figure}
    \centering
    \includegraphics[width = 0.5\textwidth]{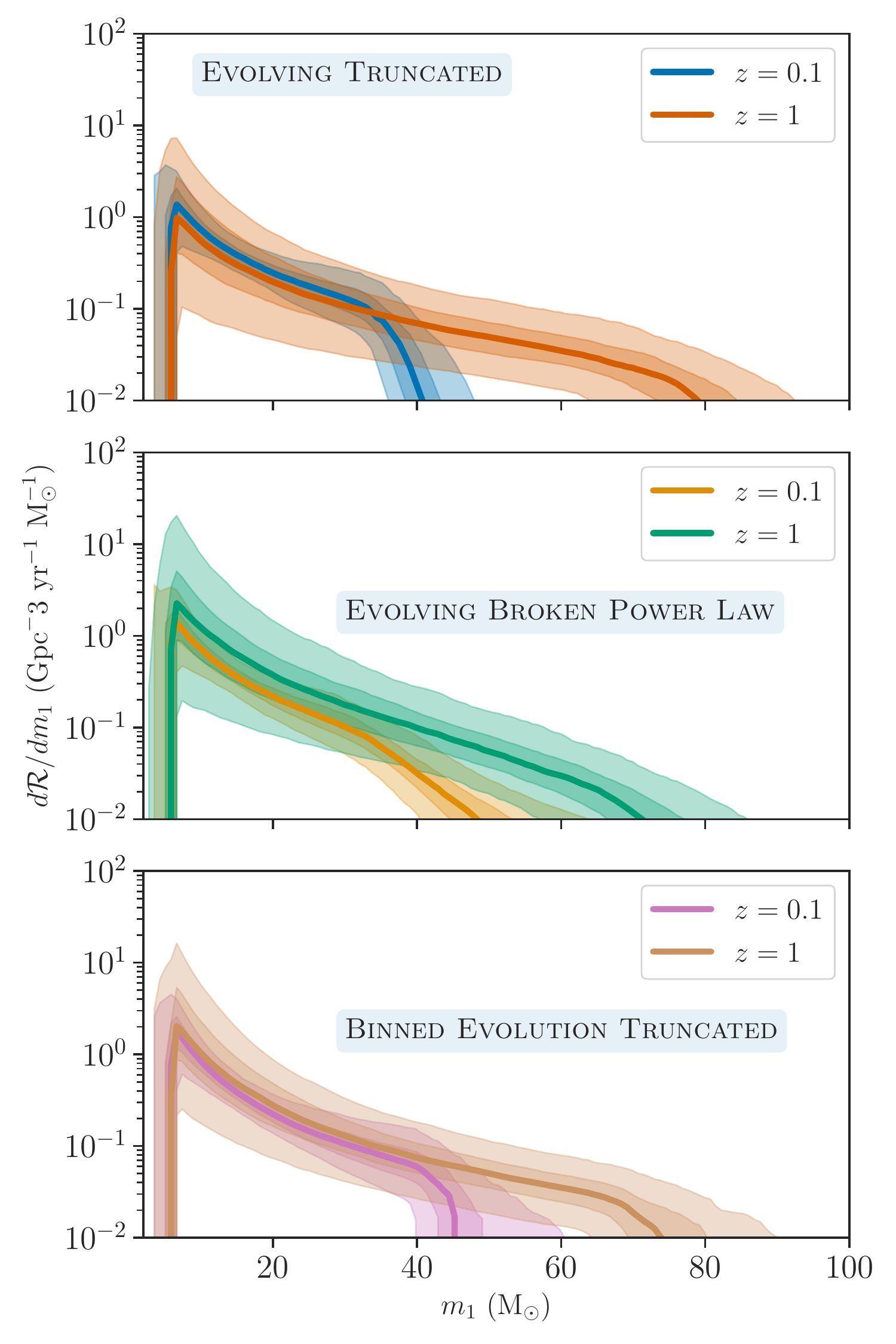}
    \caption{
        Rate density as a function of $m_1$ inferred under models that allow the mass distribution and the overall rate to evolve with redshift.
        Solid lines denote the median $d\mathcal{R}/dm_1$ at each $m_1$ and shaded regions correspond to the 50\% and 90\% symmetric credible regions.
        (\emph{top}) A sharp high-mass cutoff at $m_\mathrm{break}(z)$, fixing the high-mass ($m_1 > m_\mathrm{break}$) power law slope to $\alpha_2 = -20$.
        (\emph{middle}) Models with variable $\alpha_2$ and $m_\mathrm{break}$.
        The top and middle panels both assume that only $m_\mathrm{break}$ evolves with redshift (see Table~\ref{tab:models overview}), while the other mass distribution parameters are constant in redshift.
        (\emph{bottom}) the \binned~model, in which a separate \truncated~mass distribution is fit to systems with $z < 0.4$ and $z > 0.4$, and all mass distributions parameters are allowed to vary between the redshift bins.
    }
    \label{fig:Rm1}
\end{figure}

Fig.~\ref{fig:Rm1} shows the rate density as a function of $m_1$, $\dd\mathcal{R} / \dd m_1$, at two redshifts, $z = 0.1$ and $z = 1$, inferred under the \Etruncated~ and  \Ebrokenpowerlaw~models described above and the \binned~model of Sec.~\ref{sec:binned}.
Note that our models allow for an overall evolution of the merger rate through $f(z)$ in Eq.~\ref{eq:diff-rate-general}, and we generally infer different values for the total merger rate at $z = 1$ and $z = 0.1$.
The \Etruncated~and \binned~models (first and third panels, respectively) both assume that the primary mass distribution has a sharp maximum mass cutoff; under these models, we infer that the mass distribution extends to higher masses at $z = 1$ than at $z = 0.1$.
Meanwhile, the \Ebrokenpowerlaw~model allows for a consistent shape to the mass distribution at $z = 1$ and $z = 0.1$, although evolution towards higher masses at high redshifts is also possible.

\begin{figure}
    \centering
    \includegraphics[width = 0.5\textwidth]{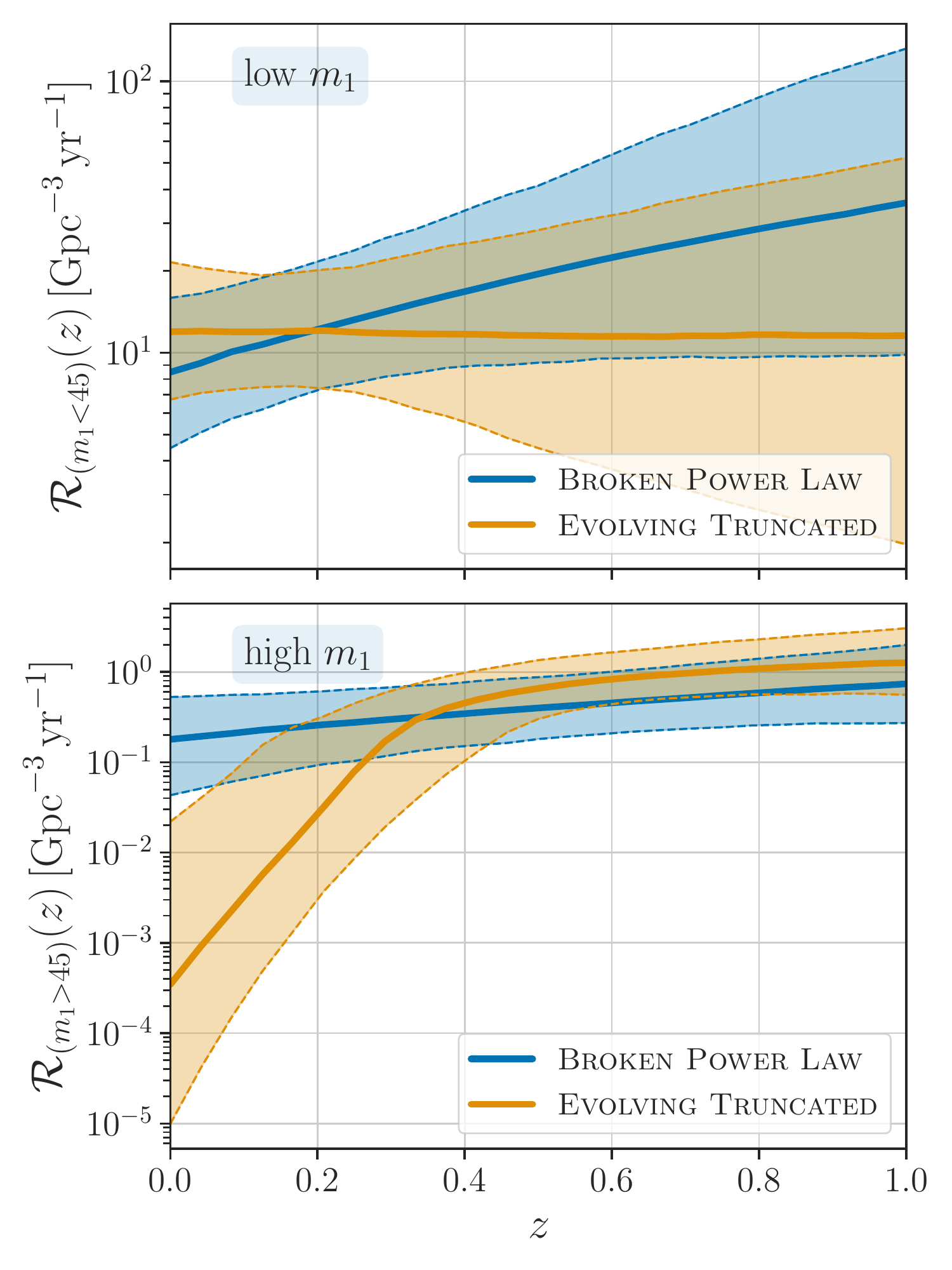}
    \caption{
        Rate evolution as a function of redshift for BBH systems with (\emph{top}) $m_1 < 45\,\msun$ and (\emph{bottom}) $m_1 > 45\,\msun$, in the non-evolving \brokenpowerlaw~model (\emph{blue}) and the \Etruncated~model (\emph{orange}).
        If we assume that the mass distribution does not evolve with redshift, the evolution of the merger rate in any given mass range follows $f(z) \propto (1 + z)^\kappa$.
        When we allow the mass distribution to evolve with redshift, the merger rate for BBH systems with $m_1 > 45\,\msun$ increases with redshift more rapidly, with a small rate $\mathcal{R}_{(m_1 > 45)}< 10^{-2}\,\mathrm{Gpc}^{-3}\,\mathrm{yr}^{-1}$ at $z = 0$ but a similar rate to the non-evolving mass distribution model at $z \gtrsim 0.3$.
        Meanwhile, for low-mass events, the evolving mass distribution predicts a slightly smaller merger rate  at high $z$ compared to the non-evolving mass distribution.
    }
    \label{fig:rate_m1above45}
\end{figure}

Another way of understanding the evolution of the mass distribution is seen in Fig.~\ref{fig:rate_m1above45}, which shows the merger rate as a function of redshift for systems with $m_1 < 45\,\msun$ (top panel) compared to $m_1 > 45\,\msun$ (bottom panel).
The blue bands show the \brokenpowerlaw~mass distribution, in which the merger rate can evolve with redshift, but the evolution is independent of the masses. With this model, we find that the overall merger rate likely evolves, with rate evolution parameter \result{$\kappa = 2.1^{+2.2}_{-1.9}$} ($\kappa = 0$ corresponds to a non-evolving rate).
The orange bands show the \Etruncated~model, in which the mass distribution, as well as the merger rate, can evolve with redshift.
This model finds that the merger rate for systems with $m_1 > 45\,\msun$ evolves significantly (orange band, bottom panel).
At low redshifts, there are few such systems, with a merger rate $\lesssim 0.01\, \mathrm{Gpc}^{-3}\,\mathrm{yr}^{-1}$ at $z = 0$, but at $z = 1$, the merger rate of $m_1 > 45\,\msun$ systems reaches $\sim10\, \mathrm{Gpc}^{-3}\,\mathrm{yr}^{-1}$. Meanwhile, as seen by the slope of the orange band in the top panel, the merger rate for systems with $m_1 < 45\,\msun$ might not evolve at all, or may even decrease with increasing redshift.
Therefore, the \Etruncated~model infers less evolution in the total merger rate, finding smaller values of \result{$\kappa = 0.0^{+2.4}_{-2.6}$}, compared to the \brokenpowerlaw~model.
The correlation between the evolution of the total merger rate and the evolution of the mass distribution is to be expected; see the discussion in Section~\ref{sec:binned}. In order to explain the number of observations at high redshifts, we infer either that the total merger rate must increase with redshift, or that the fraction of high-mass events must increase while the total merger rate stays approximately constant.

\begin{figure}
    \centering
    \includegraphics[width=0.5\textwidth]{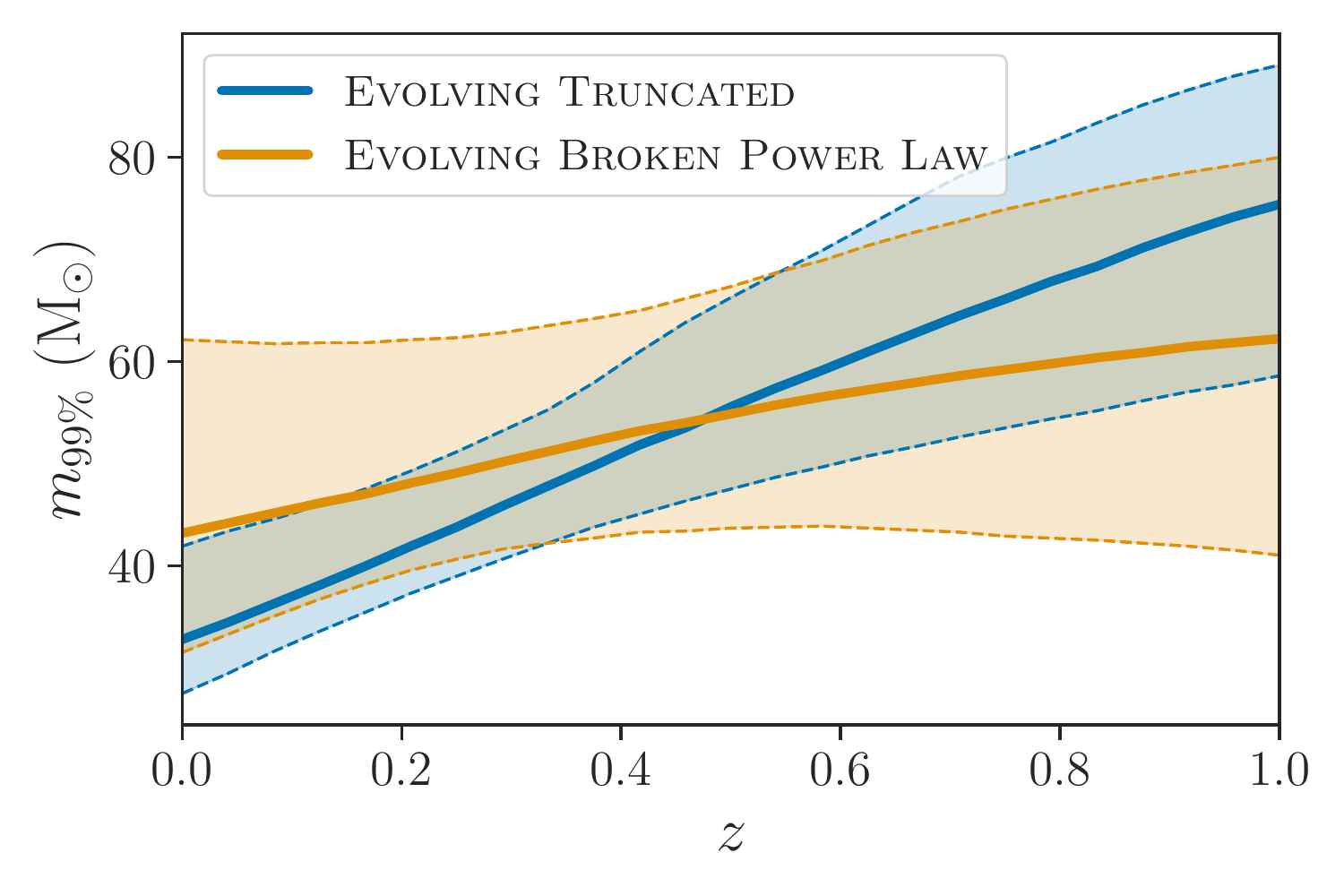}
    \caption{
        The $99^\mathrm{th}$ percentile of the primary mass distribution, $m_{99\%}$, as a function of redshift.
        Solid lines show the median and shaded bands show symmetric 90\% credible intervals.
        Assuming that the BH mass distribution has a sharp cutoff (\Etruncated~model), it likely increases with redshift (\result{$> 99.9\%$} credibility).
        Assuming an \Ebrokenpowerlaw~model, there is a weaker preference for the location of the break to evolve (\result{83\%} credibility).
    }
    \label{fig:m99vz}
\end{figure}

\begin{figure*}
    \begin{minipage}{0.5\textwidth}
    \centering
    \includegraphics[width = 1.0\textwidth]{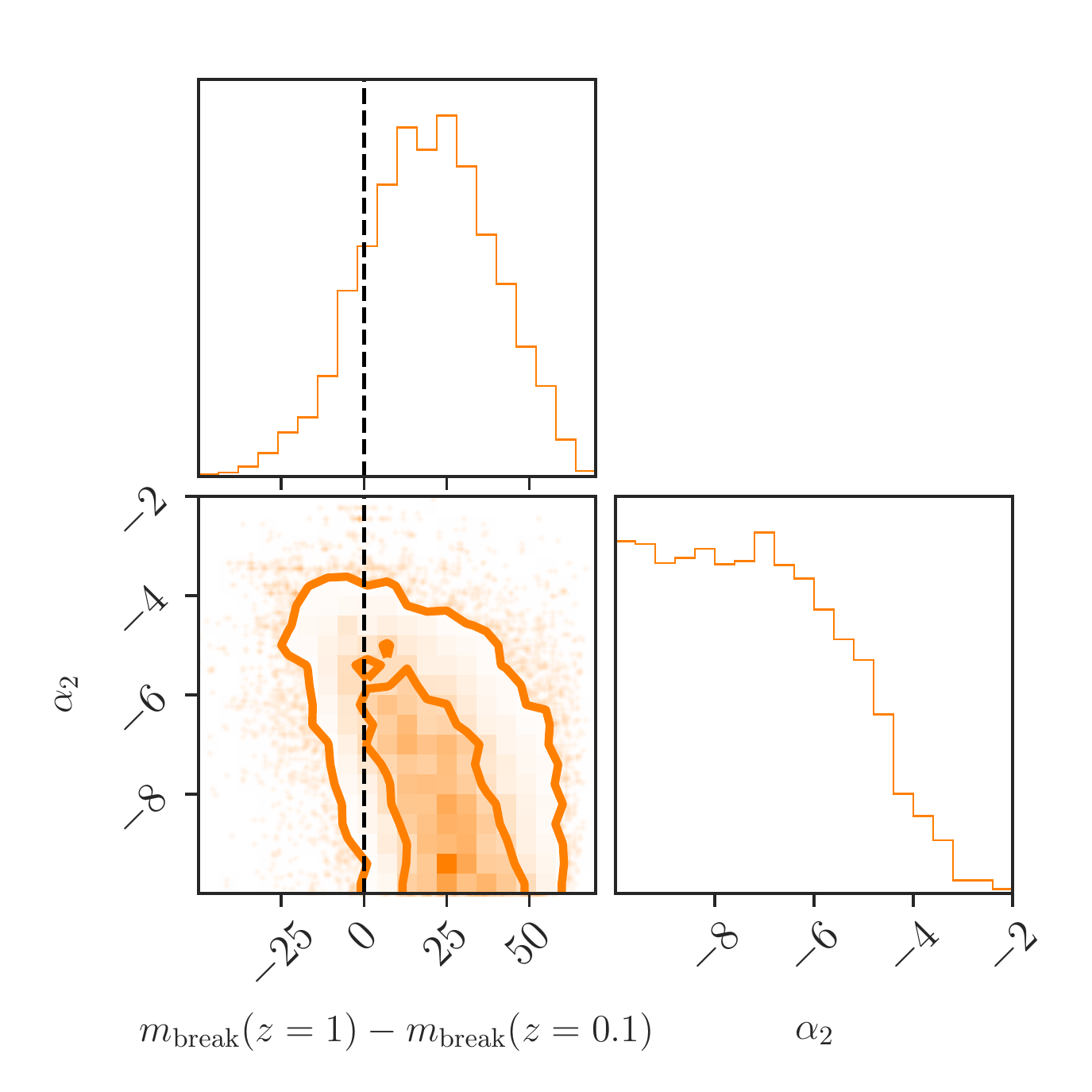}
    \caption{
        Corner plot showing the correlation between the evolution of $m_\mathrm{break}$ in the \Ebrokenpowerlaw~model and the high-mass power-law slope $\alpha_2$.
        The vertical dashed line corresponds to no evolution of the mass distribution.
        For a sharp cutoff $\alpha_2 \ll 0$, we find a strong preference for evolution, but for a shallower cutoff ($\alpha_2 \gtrsim -6$), the observed events are consistent with no evolution.
    }
    \label{fig:mbalpha2corner}
    \end{minipage}
    \begin{minipage}{0.5\textwidth}
    \centering
    \includegraphics[width = 1.0\textwidth]{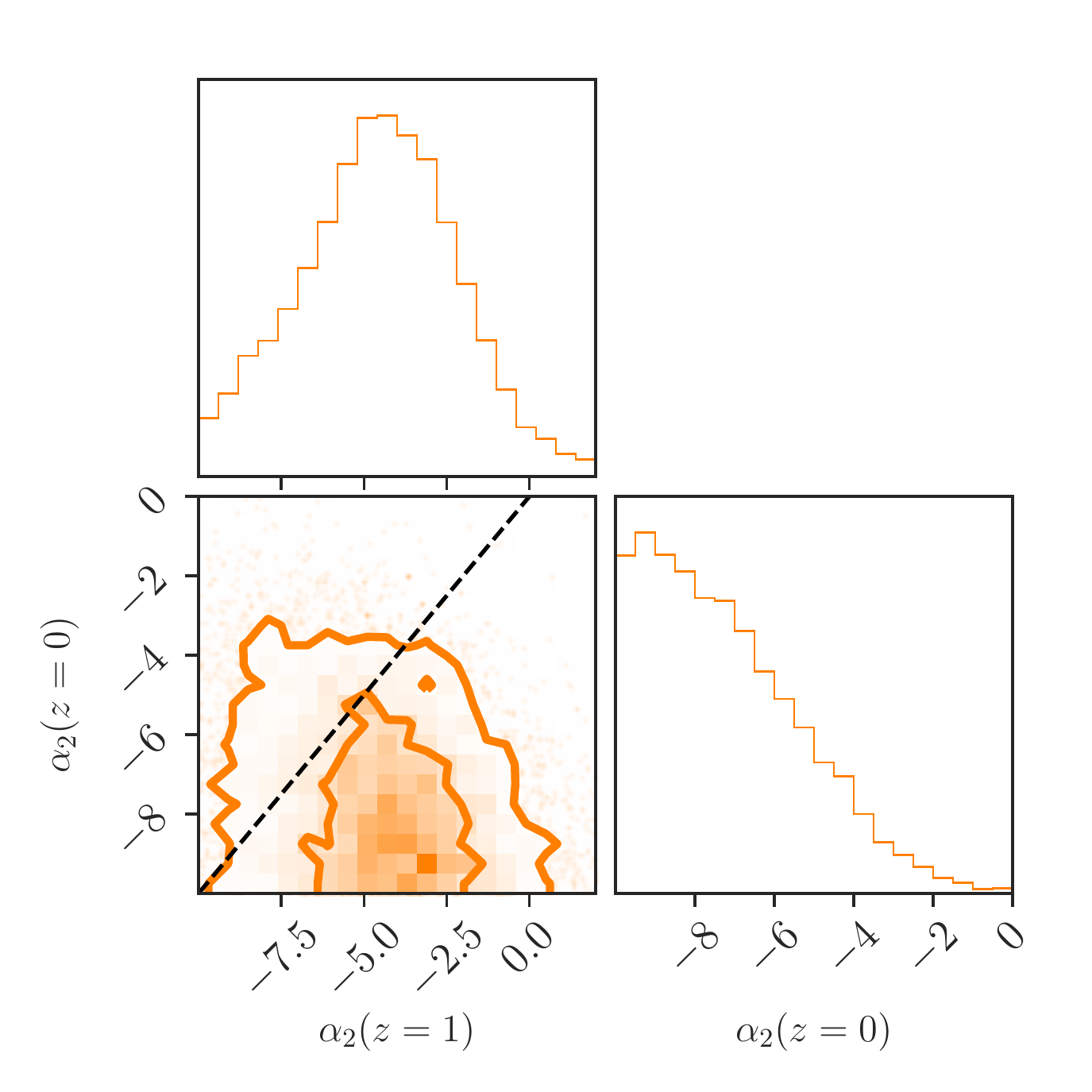}
    \caption{
        Corner plot showing the correlation between $\alpha_2(z = 1)$ and $\alpha_2(z = 0)$ inferred under an \Abrokenpowerlaw~model in which $\alpha_2$ evolves with redshift.
        The dashed black line shows $\alpha_2(z = 1) = \alpha_2(z = 0)$, corresponding to no redshift evolution.
        Similarly to the model with an evolving $m_\mathrm{break}$, when $\alpha_2(z = 0)$ is steep, we recover a preference for the mass distribution to evolve.
    }
    \label{fig:alpha2evolcorner}
    \end{minipage}
\end{figure*}

For the \Etruncated~ and \Ebrokenpowerlaw~models, we show the inferred $99^\mathrm{th}$ percentile of the $m_1$ distribution, $m_{99\%}$, as a function of redshift in Fig.~\ref{fig:m99vz}.
As we saw in Figs.~\ref{fig:Rm1} and \ref{fig:rate_m1above45}, the \Etruncated~model finds a strong preference for the high-mass cutoff to increase with increasing redshift (\result{$>99.9\%$}).
On the other hand, if the primary mass distribution follows an \Ebrokenpowerlaw~model, the preference for mass evolution is much weaker.
Marginalizing over the high-mass power law slope $\alpha_2$, we find that $m_\mathrm{break}$ increases with increasing redshift at \result{83\%} credibility.
We stress that even this mild preference for evolution in the \Ebrokenpowerlaw~model depends on the choice of $\alpha_2$ prior, because of the correlation between the steepness of $\alpha_2$ and the evolution of $m_\mathrm{break}$.
This degeneracy between the abruptness of the high-mass cutoff and the preference for evolution can be seen in Fig.~\ref{fig:mbalpha2corner}, which shows the correlation between the evolution of $m_\mathrm{break}$ and $\alpha_2$, the power-law slope above $m_\mathrm{break}$.
In the limit of large negative $\alpha_2 \lesssim -8$, the \Ebrokenpowerlaw~model approaches an \Etruncated~model, and we find a strong preference for $m_\mathrm{break}$ to evolve.
Meanwhile, for shallower values of $-6 \lesssim \alpha_2 \lesssim -4$, the data is consistent with $m_\mathrm{break}(z = 1) = m_\mathrm{break}(z = 0)$.
Our final posterior, then, depends on how much prior volume we include below $\alpha_2 \lesssim 8$.

While we have chosen to parameterize evolution in the \Ebrokenpowerlaw~model with a redshift-dependent $m_\mathrm{break}$ parameter, we find similar results when we instead consider a redshift-dependent $\alpha_2(z)$.
This is the \Abrokenpowerlaw{} described in Table~\ref{tab:models overview}.

This alternate parameterization may approximate the scenario in which the break mass $m_\mathrm{break}$ represents the lower edge of the pair instability gap (fixed across cosmic time), so that systems with $m_1 > m_\mathrm{break}$ belong to a subpopulation that contaminates the gap.
In this scenario, $\alpha_2$ sets the rate of this subpopulation, which can become more (less negative $\alpha_2$) or less (more negative $\alpha_2$) dominant at different redshifts.
Fig.~\ref{fig:alpha2evolcorner} shows the posterior on $\alpha_2(z = 0)$ and $\alpha_2(z = 1)$ under this model, marginalizing over the other model parameters.
As in Fig.~\ref{fig:mbalpha2corner}, we infer that if the $z = 0$ mass distribution has a sharp cutoff ($\alpha_2^0 \lesssim -8$), $\alpha_2$ probably becomes less negative with increasing redshift, corresponding to a higher merger rate for systems with $m_1 > m_\mathrm{break}$.
However, the data remain consistent with a mass distribution that is independent of redshift as long as $\alpha_2 \gtrsim -6$.
Additionally, we recover similar results for the $99^\mathrm{th}$ percentile of the mass distribution as a function of redshift.


\subsection{Sensitivity to GW190521}
\label{sec:GW190521}

Since GW190521 is the most massive event in our catalog, one might suspect that our conclusions regarding the high-mass end of the BBH population are driven by this event.
To test this, we repeat our analyses while excluding GW190521 from the sample, and we verify that our results are robust in these leave-one-out analyses.
If we model the mass distribution with a sharp maximum mass cutoff and exclude GW190521 from the analysis, we recover that the location of the cutoff must evolve with redshift (\result{96.4\%} credibility in the   \Etruncated{} model; \result{95.2\%} credibility in the \binned~model).
For the \Ebrokenpowerlaw~model, any preference for evolution of the mass distribution is slightly weakened with the exclusion of GW190521, and our conclusion remains that the data is consistent with a non-evolving mass distribution.


\subsection{Robustness to Sensitivity Estimate}
\label{sec:selection}

As described in Section~\ref{sec:methods}, when fitting the population models to the data, we must account for GW selection effects. For a given population model described by $\Lambda$, we estimate the detectable fraction $\xi(\Lambda)$ of Eq.~\ref{eq:xi} by performing a Monte Carlo integral over a set of injections detected by the GWTC-2 search pipelines, re-weighted by the population model at $\Lambda$ \citep{Farr2019}. To be detected, an injection must have a FAR of less than one per year, matching the FAR threshold we use when selecting events for our analysis. Crucially, if there were a systematic error in the injections' estimated FARs that was correlated in source-frame mass and redshift, the mass distribution inference could artificially prefer redshift evolution. In particular, an apparent dearth of sources with high mass and low redshift could be explained as an \textit{over}-estimation of the sensitivity to high-mass, low-redshift systems.

As a crude test of the effect of mis-estimated sensitivity, we perform multiple population inferences on O3a events with a modified \truncated\ primary-mass model that allows linear evolution of the maximum mass and power-law slope with redshift.
In each run, we throw away different fractions of found injections at high total mass ($M>100M_\odot$) and low redshift ($z<0.4$), going all the way up to throwing away 90\% of those injections.
This simulates selection functions with ever smaller low-redshift, high-mass sensitivities.
We find that tossing out injections does not eliminate the preference for mass distribution evolution with this \truncated\ model, suggesting that sensitivity mis-estimation is likely not driving our conclusions that the data prefer a mass distribution that either evolves with redshift or has additional features beyond a simple truncated power law.

Other systematic uncertainties that may lead to spurious conclusions about the evolution of the mass distribution include the possibility of strongly lensed GW events in the sample~\citep{2017PhRvD..95d4011D} or deviations from the assumed cosmology~\citep{2019ApJ...883L..42F}. We do not account for these possibilities here. The probability that our sample contains one or more strongly lensed GW events is very small, as only $\sim 1/1000$ events are expected to be lensed~\citep{2018MNRAS.476.2220L,2018MNRAS.480.3842O}. For our analyses, we assume the cosmological parameters from~\citet{2016A&A...594A..13P} for consistency with~\citet{2020arXiv201014527A}. In principle the cosmological parameters can be simultaneously inferred with the mass distribution~\citep{2019ApJ...883L..42F}.


\section{Posterior predictive checks}
\label{sec:checks}

\begin{figure*}
    \centering
    \includegraphics[width = \textwidth]{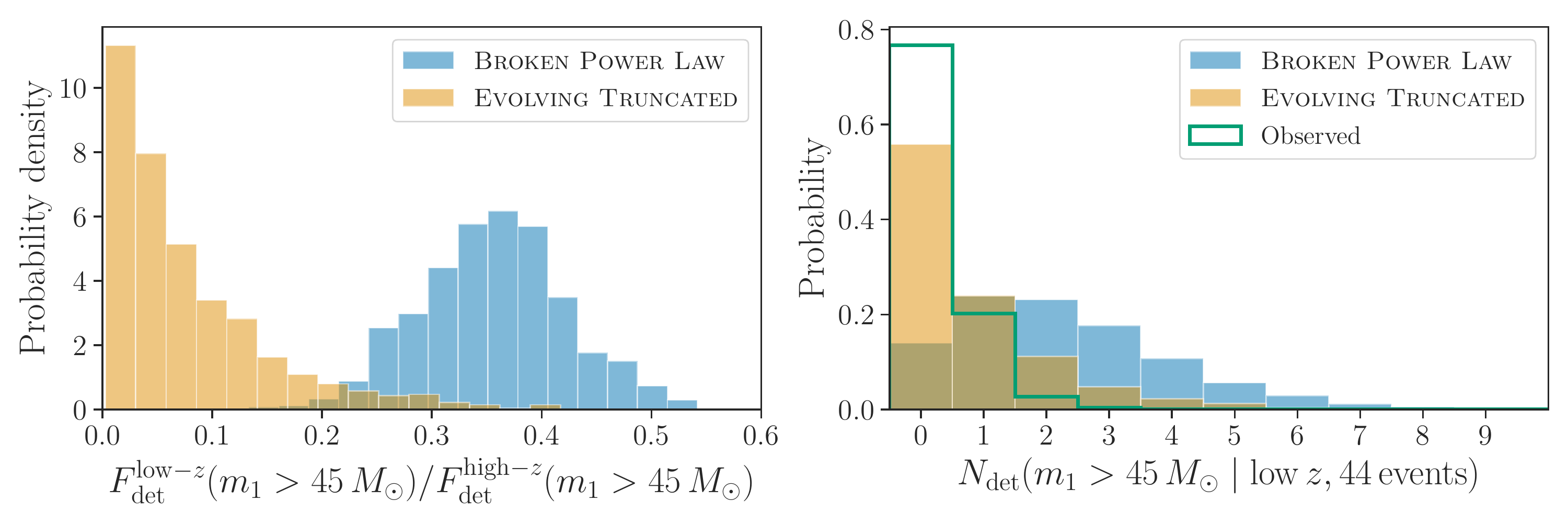}
    \caption{
        (\emph{left}) Ratio of the detection rate of high-mass events ($m_1 > 45\,\msun$) between high-redshifts and low-redshifts events (split at the median redshift).
        Model predictions refer to the true detection rate and do not account for Poisson uncertainty that arises for a finite number of observations.
        (\emph{right}) The expected number of low-redshift observations with $m_1 > 45\,\msun$ given a total of 44 events.
        The unfilled green histogram corresponds to the GWTC-2 events; the uncertainty reflects measurement uncertainty in their source parameters inferred under the \brokenpowerlaw~model.
        Both models are consistent with GWTC-2's observation of nearly zero low-redshift, $m_1 > 45\,\msun$ events.
    }
    \label{fig:Fdet_PPC}
\end{figure*}

We find that the data is consistent with two interpretations: a non-evolving \brokenpowerlaw~with a relatively shallow high-mass slope, or an \Etruncated~model.
In this section, we carry out posterior predictive checks to examine the features of the data that are most consistent with each interpretation, and discuss how future data will distinguish between the two scenarios.

As a first check, we revisit the feature highlighted in Figs.~\ref{fig:m1-z-catalog} and~\ref{fig:m1_versus_z_truncated}: the missing high-mass, low-redshift observations in GWTC-2.
Fundamentally, consistency between observations and our models boils down to the relative fractions of events detected with high masses ($m_1 \gtrsim 45\,\msun$) at low- and high-redshift.
The left panel of Fig.~\ref{fig:Fdet_PPC} shows the uncertainty in this ratio under the non-evolving \brokenpowerlaw~and \Etruncated~models.
The \brokenpowerlaw~model generally predicts a few low-redshift events with $m_1 > 45\,\msun$ while the \Etruncated~model predicts fewer, or even zero, such events.
In principle, then, we should be able to distinguish between these two models given enough observations.
That is to say, we will prefer the non-evolving \brokenpowerlaw~if we see more than one high-mass, low-redshift event for every $\sim 5$ high-mass, high-redshift event.
If we see fewer high-mass, low-redshift events, the \Etruncated~model will be preferred.

However, Poisson uncertainty with the current limited set of events is large enough that we do not strongly favor either interpretation.
The right panel of Fig.~\ref{fig:Fdet_PPC} shows the distribution of the number of detected events with $m_1>45\,\msun$ at low redshift out of 44 events under each model.
We find that even without redshift evolution, the \brokenpowerlaw~model predicts zero low-redshift, high-mass events out of 44 detections \result{15\%} of the time, and $\leq 1$ such events \result{$38\%$} of the time.
Both models, then, are consistent with the current absence of detections.

\begin{figure*}
    \begin{minipage}{0.55\textwidth}
    \centering
    \includegraphics[width = 1.0\textwidth]{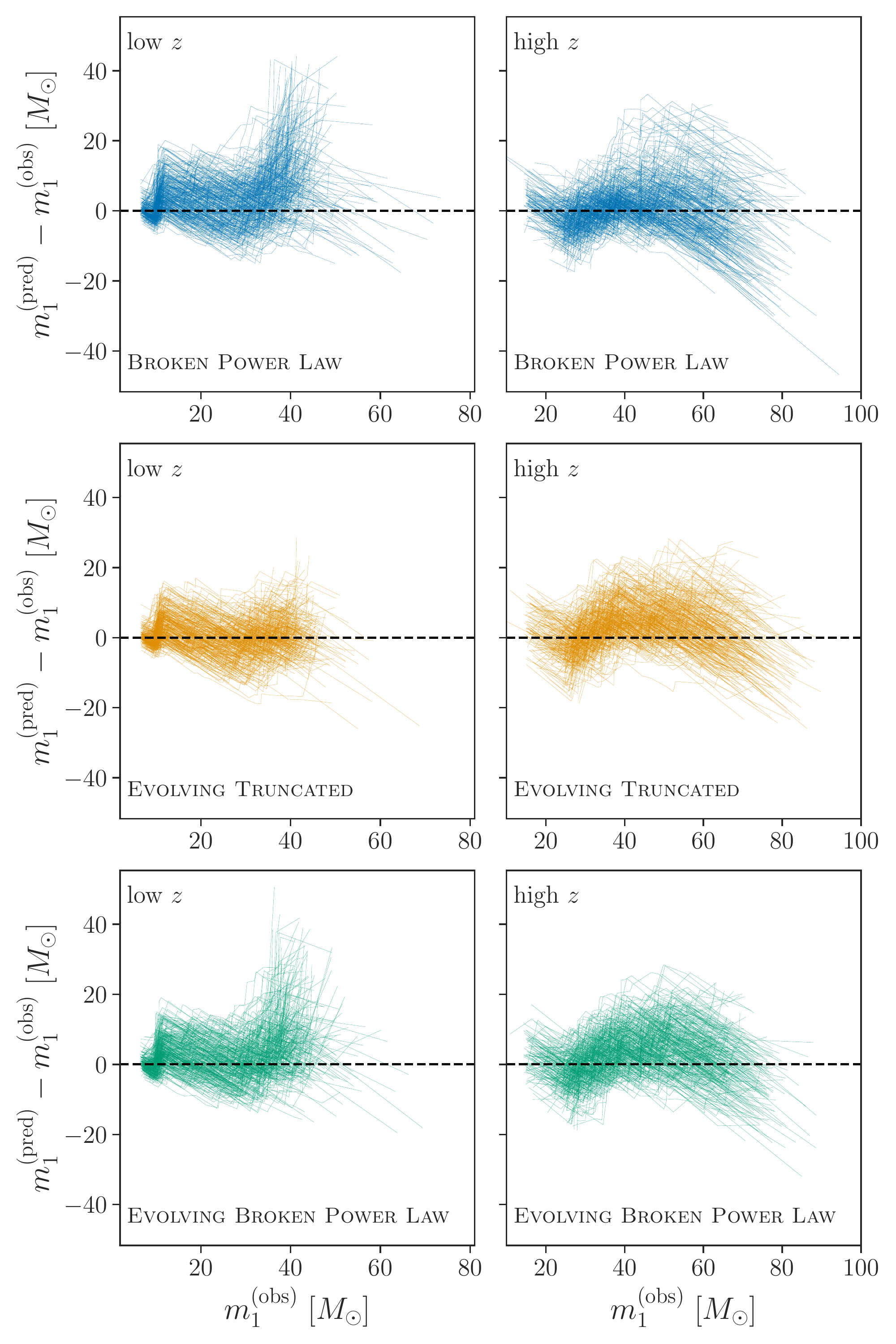}
    \caption{
        Difference between sorted sets of predicted ($m_1^{(\mathrm{pred})}$) and observed ($m_1^{\mathrm{obs})}$) primary masses as a function of the observed primary mass for (\emph{top}) \brokenpowerlaw, (\emph{middle}) \Etruncated, and (\emph{bottom}) \Ebrokenpowerlaw~models.
        Events are separated into (\emph{left}) low-redshift and (\emph{right}) high-redshift subsets.
        Each line represents a different realization of predicted and observed masses drawn from the corresponding hyperposterior.
    }
    \label{fig:lightning}
    \end{minipage}
    \begin{minipage}{0.45\textwidth}
    \centering
    \includegraphics[width = 1.0\textwidth]{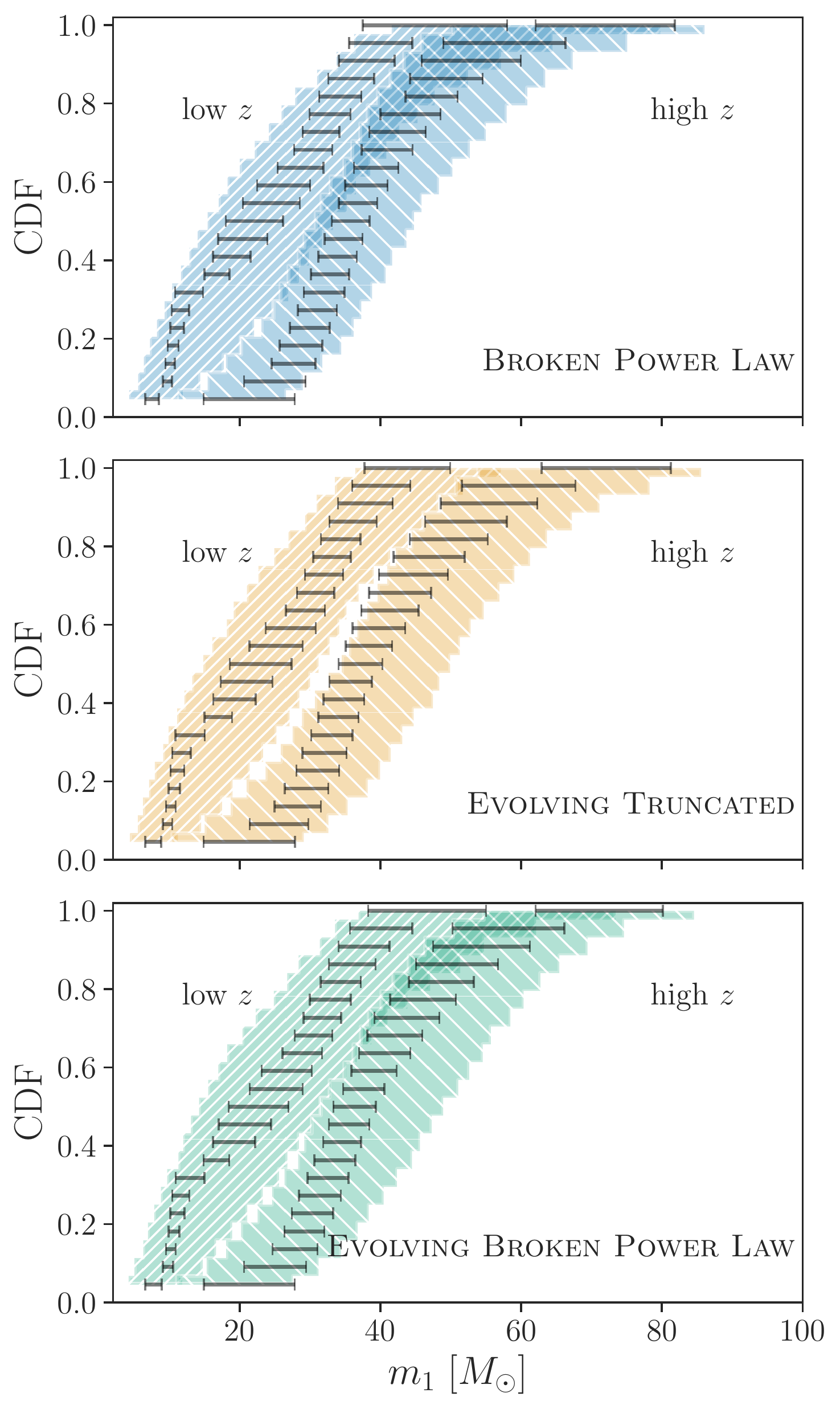}
    \caption{
        Comparison between 90\% symmetric credible regions for the predicted cumulative distribution function (shaded bands) and the empirical cumulative distribution (black bars) of primary masses for (\emph{top}) \brokenpowerlaw, (\emph{middle}) \Etruncated, and (\emph{bottom}) \Ebrokenpowerlaw~models.
        Both predicted and observed events are separated into low-redshift and high-redshift bins, split at the median redshift.
    }
    \label{fig:cdfs}
    \end{minipage}
\end{figure*}

While both evolving and non-evolving models are consistent with our current (lack of) observations of high-mass, low-redshift events, we additionally test the general goodness-of-fit of the entire mass distributions.
Figs.~\ref{fig:lightning} and~\ref{fig:cdfs} demonstrate this second check.

Specifically, Fig.~\ref{fig:lightning} demonstrates the consistency of the inferred population model with the observations by drawing 44 predicted events from the population hyperposterior, sorting them, and then comparing the predicted ($m_1^{(\mathrm{pred})}$) and observed ($m_1^{(\mathrm{obs})}$) masses as a function of $m_1^{(\mathrm{obs})}$.
As in the earlier posterior predictive checks, for every population hyperposterior sample, we draw one $m_1^{(\mathrm{obs})}$ sample per event from its population-reweighted posterior.
If the models capture the behavior of the data well, then the difference $m_1^{(\mathrm{pred})} - m_1^{(\mathrm{obs})}$ should remain near zero for all masses, regardless of the events' redshifts.
We demonstrate this by dividing the sample into low- and high-redshift bins.
The largest deviations occur for the non-evolving \brokenpowerlaw~model at low redshift (top-left panel), where the model tends to overpredict the largest observed $m_1$.
The \Etruncated~model appears to be a better fit at low-$z$ (middle-left panel), but there may be hints of a deviation at high-$z$ (middle-right panel) where the model tends to systematically overpredict the masses of the $m_1\sim 40\,\msun$ events.

Fig.~\ref{fig:cdfs} provides a complimentary view, showing the uncertainty in the predicted cumulative distribution functions (CDFs) over $m_1$, calculated from drawing sets of 44 predicted events from the population hyperposterior, along with the empirical cumulative distribution of observed $m_1$, again binned into low- and high-redshift sets.
Consistency in these plots corresponds to predicted CDF bands that encompass the black uncertainty bars from the individual events.
At low-$z$, all of the models shown are able to fit the events well, although, again, the \Etruncated~and \Ebrokenpowerlaw~models better limit the maximum predicted $m_1$ at low redshifts to the observed value, while the \brokenpowerlaw~model often overpredicts the most massive observation at low $z$.
At high-$z$, we see that the \Etruncated~model tends to predict a longer tail to high masses, as the predicted CDFs are slightly shifted to the right compared to the observed high-$z$ events.
Although the \brokenpowerlaw, \Etruncated, and \Ebrokenpowerlaw~models differ in their predictions, they all provide adequate fits to the data within the current uncertainties. 


\section{Conclusion and future prospects}
\label{sec:conclusion}

We have fit the GWTC-2 events to redshift-dependent mass models, investigating the evolution of the BBH mass distribution across cosmic time. We explored the apparent dearth of high-mass black-hole mergers at low redshift, showing that it can be explained by either a BBH mass distribution that evolves with redshift or by a mass distribution that contains features beyond a truncated power law. In a non-evolving mass distribution, these beyond-power law features must suppress the rate of high-mass systems ($m_1 > 45\,\msun$) by including, for example, a break in the power law, in agreement with the conclusions of~\citet{2020arXiv201014533T}.

We additionally confirmed that our conclusions are not driven by any particular event, showing that the results are qualitatively unchanged when we remove the heaviest system detected to date: GW190521.
At the same time, we confirmed that possible gross miss-estimation of the sensitivity of our searches, which could systematically bias our belief about our ability to detect high-mass, low-redshift events, could not account for the preference for primary mass distributions with a sharp maximum cutoff to evolve with redshift.

If a \truncated~power law with a sharp maximum mass cutoff is assumed for the primary masses of BBH systems, we find that the mass distribution must evolve between $z = 0$ and $z = 1$ with \result{$>99\%$ credibility}.
In this case, the total merger rate is consistent with binaries being uniformly distributed in comoving volume, but the types of binaries must change as a function of redshift.
If, on the other hand, the data is described by a power law with a \textit{break} rather than a sharp maximum mass cutoff, the data are consistent with a non-evolving mass distribution.
This model prefers an overall rate density that increases with increasing redshift, rather than being uniformly distributed in comoving volume.
Both models fit the current data equally well, but with additional events, we will be able to distinguish them.

\begin{figure}
    \centering
    \includegraphics[width=0.5\textwidth]{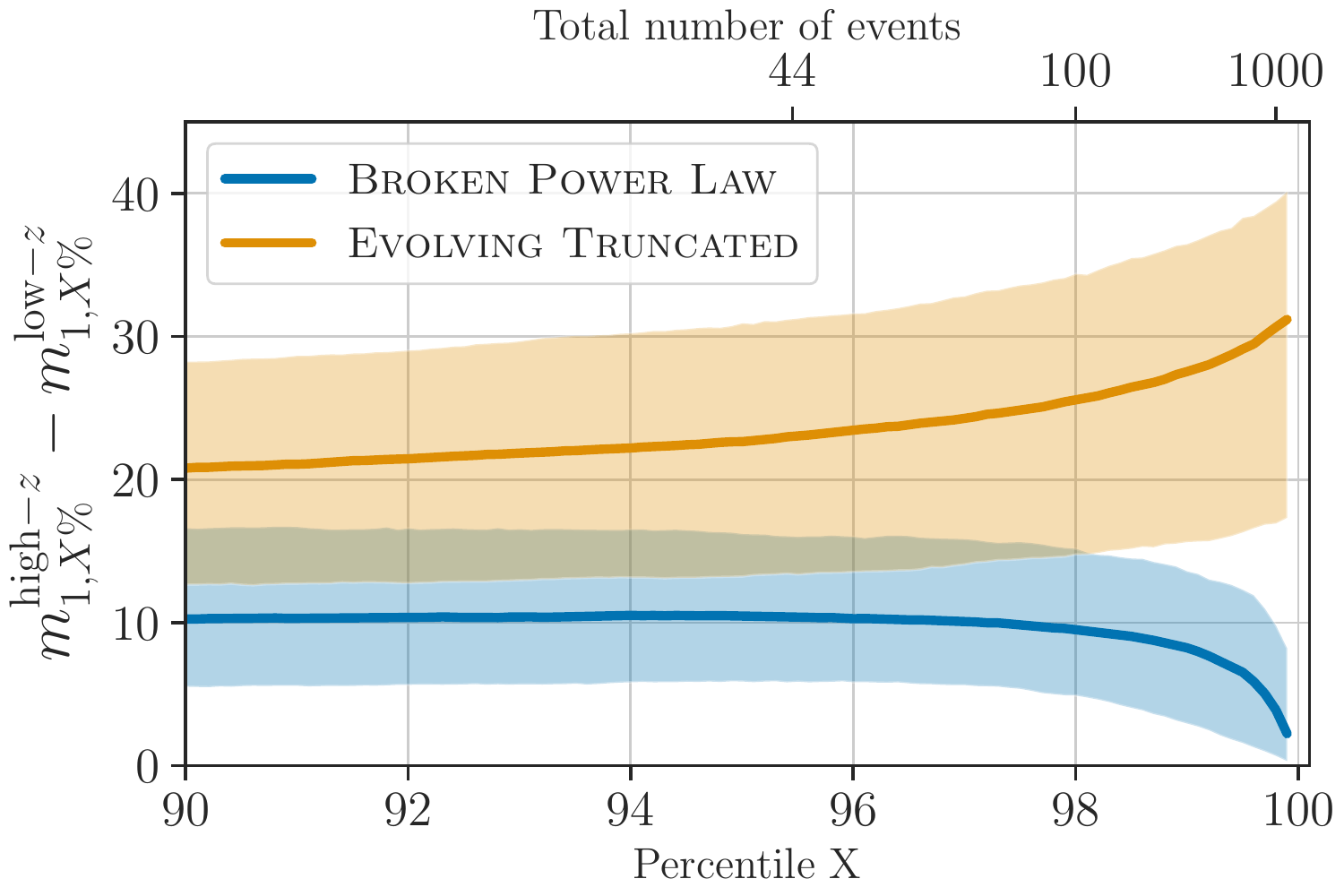}
    \caption{
        Differences in the primary mass percentiles between detected events at high and low redshift as a function of the percentile.
        Shaded regions correspond to 90\% uncertainty from the hyperposterior for the \brokenpowerlaw~(\emph{blue}) and \Etruncated~(\emph{orange}) models. The approximate total number of events needed to resolve each quantile is shown by the top x-axis.
        The current set of events only probes up to percentiles $\approx 95\%$, and more events will be needed in order to distinguish between the models.
    }
    \label{fig:diff_vs_quantile}
\end{figure}

Figure~\ref{fig:diff_vs_quantile} explores when we expect to be able to distinguish between an evolving mass distribution with a sharp maximum mass cutoff and a non-evolving mass distribution with a break in the power law, rather than a cutoff.
We plot the difference in the mass scales corresponding to the percentiles of detected events in high-redshift and low-redshift bins.
As we observe more events, we begin to resolve higher percentiles in the observed mass distribution; out of $N/2$ events in each redshift bin, we expect the most massive observed event in each bin to be at the $\sim (1 - \frac{2}{N})$ quantile.
Since we can see more massive events out to higher redshifts, even if the underlying mass distribution is the same at all redshifts, the $X\%$ primary-mass percentile in the high-redshift bin, $m_{1,X\%}^{\mathrm{high-}z}$, will correspond to a larger mass than the same percentile in the low-redshift bin, $m_{1,X\%}^{\mathrm{low-}z}$. Therefore, for a non-evolving mass distribution, or one that favors larger masses at higher redshifts, the difference $m_{1,X\%}^{\mathrm{high-}z} - m_{1,X\%}^{\mathrm{low-}z}$ is positive.
Shaded regions correspond to the uncertainty from different fair draws from the hyperposterior for each model.
As we start probing higher percentiles of the observed mass distribution, if the mass distribution does not evolve with redshift (\brokenpowerlaw~model, in blue), the heaviest events observed in the low redshift bin approach the masses of the heaviest events in the high redshift bin, and so the difference decreases.
However, if the mass distribution evolves with redshift (\Etruncated~model, in orange), the underlying \textit{astrophysical} mass distribution skews to higher masses at higher redshifts, and so the difference in masses among the \textit{observed} high- and low-redshift events is larger than in the non-evolving case, and only increases as we consider higher percentiles of the observed distribution.
Within current statistical uncertainties, the \brokenpowerlaw~and \Etruncated~models are consistent for small percentiles ($< 95^\mathrm{th}$ percentile) but diverge at higher percentiles.
The current data set only probes approximately the $1 - \frac{2}{44} \approx 0.95$ quantile, as we have $N = 44$ total events, or two redshift bins that each contain 22 events.
Therefore, we cannot resolve the discrepancy between the models that only appears at higher percentiles with the current sample size.
However, we will probe higher percentiles as we detect more events, and may be able to confidently distinguish between these models when we obtain a factor of $\gtrsim 2$ more events.

Future GW events will allow us to not only better resolve the BBH mass distribution, they will also probe the BBH population over a much higher redshift range.
In addition to the expected increases in sensitivity for current detectors \citep{2018LRR....21....3A}, the next generation of gravitational-wave detectors will be able to probe the redshift evolution of the BBH population out to $z\sim30$~\citep{2019CQGra..36v5002H}, offering us a deeper look into the formation environments of binary black hole systems \citep{2019ApJ...886L...1V, Ng:2020qpk,2020arXiv201114541R}.

Characterizing features in the BBH mass distribution, such as a sharp cutoff, a break or a peak, and tracing their evolution with redshift provides insights into the physics of BBH formation and merger. 
Future data will allow us to distinguish between several possible scenarios.
One possibility is that the maximum core mass for pair-instability supernova (PISN) is highly sensitive to metallicity, creating a redshift-dependent cutoff in the BH mass distribution.
It may also be that multiple binary formation channels contribute differently over the age of the universe such that higher mass mergers are favored at earlier times.
For example, some combination of stellar evolution, PISN physics, and/or hierarchical or stellar mergers may produce a small tail of BBHs with primary masses $\gtrsim 45\,\msun$, and these processes may be more common at higher redshift. 
Alternatively, there may be no redshift evolution of the mass distribution at $0 < z \lesssim 1$, and instead the overall rate of mergers increases with redshift independently of the component masses.
In this case, it must be that the PISN feature in the BBH mass distribution is not a sharp cutoff, or that the processes that contaminate the PISN mass gap and give rise to systems above $m_1 \gtrsim 45\,\msun$ operate at similar relative rates throughout the observable redshift range.

The first three observing runs of the advanced LIGO and Virgo interferometers have already illuminated many new aspects of compact objects within our universe.
The first detections demonstrated \emph{what} the most prevalent detectable GW sources are: they are BBH mergers.
Subsequent studies have then asked \emph{where} the biggest black holes are, motivated by the lack of observations of massive BBH during the first two LIGO/Virgo observing runs \citep{2017ApJ...851L..25F}.
However, we can now phrase the question more precisely as \emph{when} did the largest BBH merge in the history of the Universe, which will ultimately help us determine \emph{how} black holes form and \emph{why} they merge.

\acknowledgments
We thank Daniel Wysocki for providing helpful comments on the manuscript.
M.~F. is supported by NASA through NASA Hubble Fellowship grant HST-HF2-51455.001-A awarded by the Space Telescope Science Institute.
R.~E. and D.~E.~H. are supported at the University of Chicago by the Kavli Institute for Cosmological Physics through an endowment from the Kavli Foundation and its founder Fred Kavli.
D.~E.~H. is also supported by NSF grants PHY-1708081 and PHY-2011997, and gratefully acknowledges the Marion and Stuart Rice Award.
Z.~D.~was in-part supported by NSF Graduate
Research Fellowship grant DGE-1144082.
B.~E. and B.~F. are supported by NSF grant PHY-1807046.
The  Flatiron  Institute  is  supported  by  the  Simons Foundation.
The authors are grateful for computational resources provided by the LIGO Laboratory and supported by National Science Foundation Grants PHY-0757058 and PHY-0823459.
This research was conducted in part at the Kavli Institute for Theoretical Physics at the University of California, Santa Barbara and was supported in part by the National Science Foundation under Grant No. NSF PHY-1748958.
This research has made use of data, software and/or web tools obtained from the Gravitational Wave Open Science Center (\url{https://www.gw-openscience.org/ }), a service of LIGO Laboratory, the LIGO Scientific Collaboration and the Virgo Collaboration.
\software{
\textsc{Astropy}~\citep{2018AJ....156..123A},
\textsc{corner}~\citep{2016JOSS....1...24F},
\textsc{emcee}~\citep{2013ascl.soft03002F},
\textsc{NumPy}~\citep{harris2020array},
\textsc{Matplotlib}~\citep{Hunter:2007},
\textsc{PESummary}~\citep{Hoy:2020vys},
\textsc{PyMC3}~\citep{2016ascl.soft10016S},
\textsc{PyStan}~\citep{2017JSS....76....1C},
\textsc{SciPy}~\citep{2020SciPy-NMeth},
\textsc{seaborn}~\citep{waskom2020seaborn},
\textsc{Theano}~\citep{2016arXiv160502688short}
}

\bibliography{references}{}

\begin{thebibliography}{}
\expandafter\ifx\csname natexlab\endcsname\relax\def\natexlab#1{#1}\fi
\providecommand{\url}[1]{\href{#1}{#1}}
\providecommand{\dodoi}[1]{doi:~\href{http://doi.org/#1}{\nolinkurl{#1}}}
\providecommand{\doeprint}[1]{\href{http://ascl.net/#1}{\nolinkurl{http://ascl.net/#1}}}
\providecommand{\doarXiv}[1]{\href{https://arxiv.org/abs/#1}{\nolinkurl{https://arxiv.org/abs/#1}}}

\bibitem[{Abbott {et~al.}(2018)}]{Aasi:2013wya}
Abbott, B., {et~al.} 2018, Living Rev. Rel., 21, 3,
  \dodoi{10.1007/s41114-018-0012-9}

\bibitem[{{Abbott} {et~al.}(2018){Abbott}, {Abbott}, {Abbott}, {Abernathy},
  {Acernese}, {Ackley}, {Adams}, {Adams}, {Addesso}, {Adhikari}, \&
  et~al.}]{2018LRR....21....3A}
{Abbott}, B.~P., {Abbott}, R., {Abbott}, T.~D., {et~al.} 2018, Living Reviews
  in Relativity, 21, 3, \dodoi{10.1007/s41114-018-0012-9}

\bibitem[{{Abbott} {et~al.}(2019{\natexlab{a}}){Abbott}, {Abbott}, {Abbott},
  {Abraham}, {Acernese}, {Ackley}, {Adams}, {Adhikari}, {Adya}, {Affeldt}, \&
  et~al.}]{2019PhRvX...9c1040A}
---. 2019{\natexlab{a}}, Physical Review X, 9, 031040,
  \dodoi{10.1103/PhysRevX.9.031040}

\bibitem[{{Abbott} {et~al.}(2019{\natexlab{b}}){Abbott}, {Abbott}, {Abbott},
  {Abraham}, {Acernese}, {Ackley}, {Adams}, {Adhikari}, {Adya}, {Affeldt}, \&
  et~al.}]{2019ApJ...882L..24A}
---. 2019{\natexlab{b}}, \apjl, 882, L24, \dodoi{10.3847/2041-8213/ab3800}

\bibitem[{{Abbott} {et~al.}(2019{\natexlab{c}}){Abbott}, {Abbott}, {Abraham},
  {Acernese}, {Ackley}, {Adams}, {Adhikari}, {Adya}, \&
  et~al.}]{2019arXiv191211716T}
{Abbott}, R., {Abbott}, T.~D., {Abraham}, S., {et~al.} 2019{\natexlab{c}},
  arXiv e-prints, arXiv:1912.11716.
\newblock \doarXiv{1912.11716}

\bibitem[{{Abbott} {et~al.}(2020{\natexlab{a}}){Abbott}, {Abbott}, {Abraham},
  {Acernese}, {Ackley}, {Adams}, {Adams}, {Adhikari}, {Adya}, {Affeldt}, \&
  et~al.}]{2020arXiv201014527A}
---. 2020{\natexlab{a}}, arXiv e-prints, arXiv:2010.14527.
\newblock \doarXiv{2010.14527}

\bibitem[{{Abbott} {et~al.}(2020{\natexlab{b}}){Abbott}, {Abbott}, {Abraham},
  {Acernese}, {Ackley}, {Adams}, {Adams}, {Adhikari}, \&
  et~al.}]{2020arXiv201014533T}
---. 2020{\natexlab{b}}, arXiv e-prints, arXiv:2010.14533.
\newblock \doarXiv{2010.14533}

\bibitem[{{Abbott} {et~al.}(2020{\natexlab{c}}){Abbott}, {Abbott}, {Abraham},
  {Acernese}, {Ackley}, {Adams}, {Adams}, {Adhikari}, \& et~al.}]{O3a_pe}
---. 2020{\natexlab{c}}.
\newblock \url{https://dcc.ligo.org/LIGO-P2000223/public}

\bibitem[{{Acernese} {et~al.}(2015){Acernese}, {Agathos}, {Agatsuma}, {Aisa},
  {Allemandou}, {Allocca}, {Amarni}, {Astone}, {Balestri}, {Ballardin}, \&
  et~al.}]{2015CQGra..32b4001A}
{Acernese}, F., {Agathos}, M., {Agatsuma}, K., {et~al.} 2015, Classical and
  Quantum Gravity, 32, 024001, \dodoi{10.1088/0264-9381/32/2/024001}

\bibitem[{{Astropy Collaboration} {et~al.}(2018){Astropy Collaboration},
  {Price-Whelan}, {Sip{\H{o}}cz}, {G{\"u}nther}, {Lim}, {Crawford}, {Conseil},
  {Shupe}, {Craig}, {Dencheva}, \& et~al.}]{2018AJ....156..123A}
{Astropy Collaboration}, {Price-Whelan}, A.~M., {Sip{\H{o}}cz}, B.~M., {et~al.}
  2018, \aj, 156, 123, \dodoi{10.3847/1538-3881/aabc4f}

\bibitem[{{Belczynski} {et~al.}(2010){Belczynski}, {Dominik}, {Bulik},
  {O'Shaughnessy}, {Fryer}, \& {Holz}}]{2010ApJ...715L.138B}
{Belczynski}, K., {Dominik}, M., {Bulik}, T., {et~al.} 2010, \apjl, 715, L138,
  \dodoi{10.1088/2041-8205/715/2/L138}

\bibitem[{{Brott} {et~al.}(2011){Brott}, {de Mink}, {Cantiello}, {Langer}, {de
  Koter}, {Evans}, {Hunter}, {Trundle}, \& {Vink}}]{2011A&A...530A.115B}
{Brott}, I., {de Mink}, S.~E., {Cantiello}, M., {et~al.} 2011, \aap, 530, A115,
  \dodoi{10.1051/0004-6361/201016113}

\bibitem[{{Callister} {et~al.}(2020){Callister}, {Fishbach}, {Holz}, \&
  {Farr}}]{2020ApJ...896L..32C}
{Callister}, T., {Fishbach}, M., {Holz}, D.~E., \& {Farr}, W.~M. 2020, \apjl,
  896, L32, \dodoi{10.3847/2041-8213/ab9743}

\bibitem[{{Carpenter} {et~al.}(2017){Carpenter}, {Gelman}, {Hoffman}, {Lee},
  {Goodrich}, {Betancourt}, {Brubaker}, {Guo}, {Li}, \&
  {Riddell}}]{2017JSS....76....1C}
{Carpenter}, B., {Gelman}, A., {Hoffman}, M.~D., {et~al.} 2017, Journal of
  Statistical Software, 76, 1

\bibitem[{Chen {et~al.}(2017)Chen, Holz, Miller, Evans, Vitale, \&
  Creighton}]{Chen:2017wpg}
Chen, H.-Y., Holz, D.~E., Miller, J., {et~al.} 2017.
\newblock \doarXiv{1709.08079}

\bibitem[{{Dai} {et~al.}(2017){Dai}, {Venumadhav}, \&
  {Sigurdson}}]{2017PhRvD..95d4011D}
{Dai}, L., {Venumadhav}, T., \& {Sigurdson}, K. 2017, \prd, 95, 044011,
  \dodoi{10.1103/PhysRevD.95.044011}

\bibitem[{{Dominik} {et~al.}(2015){Dominik}, {Berti}, {O'Shaughnessy},
  {Mandel}, {Belczynski}, {Fryer}, {Holz}, {Bulik}, \&
  {Pannarale}}]{2015ApJ...806..263D}
{Dominik}, M., {Berti}, E., {O'Shaughnessy}, R., {et~al.} 2015, \apj, 806, 263,
  \dodoi{10.1088/0004-637X/806/2/263}

\bibitem[{{El-Badry} {et~al.}(2019){El-Badry}, {Quataert}, {Weisz}, {Choksi},
  \& {Boylan-Kolchin}}]{2019MNRAS.482.4528E}
{El-Badry}, K., {Quataert}, E., {Weisz}, D.~R., {Choksi}, N., \&
  {Boylan-Kolchin}, M. 2019, \mnras, 482, 4528, \dodoi{10.1093/mnras/sty3007}

\bibitem[{Farr(2019)}]{Farr2019}
Farr, W.~M. 2019, Research Notes of the {AAS}, 3, 66,
  \dodoi{10.3847/2515-5172/ab1d5f}

\bibitem[{{Farr} {et~al.}(2019){Farr}, {Fishbach}, {Ye}, \&
  {Holz}}]{2019ApJ...883L..42F}
{Farr}, W.~M., {Fishbach}, M., {Ye}, J., \& {Holz}, D.~E. 2019, \apjl, 883,
  L42, \dodoi{10.3847/2041-8213/ab4284}

\bibitem[{{Farrell} {et~al.}(2020){Farrell}, {Groh}, {Hirschi}, {Murphy},
  {Kaiser}, {Ekstr{\"o}m}, {Georgy}, \& {Meynet}}]{2020arXiv200906585F}
{Farrell}, E.~J., {Groh}, J.~H., {Hirschi}, R., {et~al.} 2020, arXiv e-prints,
  arXiv:2009.06585.
\newblock \doarXiv{2009.06585}

\bibitem[{{Fishbach} \& {Holz}(2017)}]{2017ApJ...851L..25F}
{Fishbach}, M., \& {Holz}, D.~E. 2017, \apjl, 851, L25,
  \dodoi{10.3847/2041-8213/aa9bf6}

\bibitem[{{Fishbach} {et~al.}(2018){Fishbach}, {Holz}, \&
  {Farr}}]{2018ApJ...863L..41F}
{Fishbach}, M., {Holz}, D.~E., \& {Farr}, W.~M. 2018, \apjl, 863, L41,
  \dodoi{10.3847/2041-8213/aad800}

\bibitem[{{Foreman-Mackey}(2016)}]{2016JOSS....1...24F}
{Foreman-Mackey}, D. 2016, The Journal of Open Source Software, 1, 24,
  \dodoi{10.21105/joss.00024}

\bibitem[{{Foreman-Mackey} {et~al.}(2013){Foreman-Mackey}, {Conley},
  {Meierjurgen Farr}, {Hogg}, {Lang}, {Marshall}, {Price-Whelan}, {Sanders}, \&
  {Zuntz}}]{2013ascl.soft03002F}
{Foreman-Mackey}, D., {Conley}, A., {Meierjurgen Farr}, W., {et~al.} 2013,
  {emcee: The MCMC Hammer}.
\newblock \doeprint{1303.002}

\bibitem[{{Fryer} {et~al.}(2012){Fryer}, {Belczynski}, {Wiktorowicz},
  {Dominik}, {Kalogera}, \& {Holz}}]{2012ApJ...749...91F}
{Fryer}, C.~L., {Belczynski}, K., {Wiktorowicz}, G., {et~al.} 2012, \apj, 749,
  91, \dodoi{10.1088/0004-637X/749/1/91}

\bibitem[{{Hall} \& {Evans}(2019)}]{2019CQGra..36v5002H}
{Hall}, E.~D., \& {Evans}, M. 2019, Classical and Quantum Gravity, 36, 225002,
  \dodoi{10.1088/1361-6382/ab41d6}

\bibitem[{Harris {et~al.}(2020)Harris, Millman, van~der Walt, Gommers,
  Virtanen, Cournapeau, Wieser, Taylor, Berg, Smith, Kern, Picus, Hoyer, van
  Kerkwijk, Brett, Haldane, del R{'{\i}}o, Wiebe, Peterson,
  G{'{e}}rard-Marchant, Sheppard, Reddy, Weckesser, Abbasi, Gohlke, \&
  Oliphant}]{harris2020array}
Harris, C.~R., Millman, K.~J., van~der Walt, S.~J., {et~al.} 2020, Nature, 585,
  357, \dodoi{10.1038/s41586-020-2649-2}

\bibitem[{{Hogg}(1999)}]{1999astro.ph..5116H}
{Hogg}, D.~W. 1999, arXiv e-prints, astro.
\newblock \doarXiv{astro-ph/9905116}

\bibitem[{Hoy \& Raymond(2020)}]{Hoy:2020vys}
Hoy, C., \& Raymond, V. 2020.
\newblock \doarXiv{2006.06639}

\bibitem[{Hunter(2007)}]{Hunter:2007}
Hunter, J.~D. 2007, Computing in Science \& Engineering, 9, 90,
  \dodoi{10.1109/MCSE.2007.55}

\bibitem[{{Kinugawa} {et~al.}(2020){Kinugawa}, {Nakamura}, \&
  {Nakano}}]{2020arXiv200906922K}
{Kinugawa}, T., {Nakamura}, T., \& {Nakano}, H. 2020, arXiv e-prints,
  arXiv:2009.06922.
\newblock \doarXiv{2009.06922}

\bibitem[{{Kudritzki} \& {Puls}(2000)}]{2000ARA&A..38..613K}
{Kudritzki}, R.-P., \& {Puls}, J. 2000, \araa, 38, 613,
  \dodoi{10.1146/annurev.astro.38.1.613}

\bibitem[{{Kushnir} {et~al.}(2016){Kushnir}, {Zaldarriaga}, {Kollmeier}, \&
  {Waldman}}]{2016MNRAS.462..844K}
{Kushnir}, D., {Zaldarriaga}, M., {Kollmeier}, J.~A., \& {Waldman}, R. 2016,
  \mnras, 462, 844, \dodoi{10.1093/mnras/stw1684}

\bibitem[{{Li} {et~al.}(2018){Li}, {Mao}, {Zhao}, \&
  {Lu}}]{2018MNRAS.476.2220L}
{Li}, S.-S., {Mao}, S., {Zhao}, Y., \& {Lu}, Y. 2018, \mnras, 476, 2220,
  \dodoi{10.1093/mnras/sty411}

\bibitem[{{LIGO Scientific Collaboration} {et~al.}(2015){LIGO Scientific
  Collaboration}, {Aasi}, {Abbott}, {Abbott}, {Abbott}, {Abernathy}, {Ackley},
  {Adams}, {Adams}, {Addesso}, \& et~al.}]{2015CQGra..32g4001L}
{LIGO Scientific Collaboration}, {Aasi}, J., {Abbott}, B.~P., {et~al.} 2015,
  Classical and Quantum Gravity, 32, 074001,
  \dodoi{10.1088/0264-9381/32/7/074001}

\bibitem[{{Loredo}(2004)}]{2004AIPC..735..195L}
{Loredo}, T.~J. 2004, in American Institute of Physics Conference Series, Vol.
  735, Bayesian Inference and Maximum Entropy Methods in Science and
  Engineering: 24th International Workshop on Bayesian Inference and Maximum
  Entropy Methods in Science and Engineering, ed. R.~{Fischer}, R.~{Preuss}, \&
  U.~V. {Toussaint}, 195--206, \dodoi{10.1063/1.1835214}

\bibitem[{{Mandel}(2010)}]{2010PhRvD..81h4029M}
{Mandel}, I. 2010, \prd, 81, 084029, \dodoi{10.1103/PhysRevD.81.084029}

\bibitem[{{Mandel} {et~al.}(2019){Mandel}, {Farr}, \&
  {Gair}}]{2019MNRAS.486.1086M}
{Mandel}, I., {Farr}, W.~M., \& {Gair}, J.~R. 2019, \mnras, 486, 1086,
  \dodoi{10.1093/mnras/stz896}

\bibitem[{{Mapelli} {et~al.}(2019){Mapelli}, {Giacobbo}, {Santoliquido}, \&
  {Artale}}]{2019MNRAS.487....2M}
{Mapelli}, M., {Giacobbo}, N., {Santoliquido}, F., \& {Artale}, M.~C. 2019,
  \mnras, 487, 2, \dodoi{10.1093/mnras/stz1150}

\bibitem[{{Miller} {et~al.}(2020){Miller}, {Callister}, \&
  {Farr}}]{2020ApJ...895..128M}
{Miller}, S., {Callister}, T.~A., \& {Farr}, W.~M. 2020, \apj, 895, 128,
  \dodoi{10.3847/1538-4357/ab80c0}

\bibitem[{{Neijssel} {et~al.}(2019){Neijssel}, {Vigna-G{\'o}mez}, {Stevenson},
  {Barrett}, {Gaebel}, {Broekgaarden}, {de Mink}, {Sz{\'e}csi}, {Vinciguerra},
  \& {Mandel}}]{2019MNRAS.490.3740N}
{Neijssel}, C.~J., {Vigna-G{\'o}mez}, A., {Stevenson}, S., {et~al.} 2019,
  \mnras, 490, 3740, \dodoi{10.1093/mnras/stz2840}

\bibitem[{Ng {et~al.}(2020)Ng, Vitale, Farr, \& Rodriguez}]{Ng:2020qpk}
Ng, K.~K., Vitale, S., Farr, W.~M., \& Rodriguez, C.~L. 2020.
\newblock \doarXiv{2012.09876}

\bibitem[{{Ng} {et~al.}(2018){Ng}, {Vitale}, {Zimmerman}, {Chatziioannou},
  {Gerosa}, \& {Haster}}]{2018PhRvD..98h3007N}
{Ng}, K. K.~Y., {Vitale}, S., {Zimmerman}, A., {et~al.} 2018, \prd, 98, 083007,
  \dodoi{10.1103/PhysRevD.98.083007}

\bibitem[{{Oguri}(2018)}]{2018MNRAS.480.3842O}
{Oguri}, M. 2018, \mnras, 480, 3842, \dodoi{10.1093/mnras/sty2145}

\bibitem[{{Planck Collaboration} {et~al.}(2016){Planck Collaboration}, {Ade},
  {Aghanim}, {Arnaud}, {Ashdown}, {Aumont}, {Baccigalupi}, {Banday},
  {Barreiro}, {Bartlett}, \& et~al.}]{2016A&A...594A..13P}
{Planck Collaboration}, {Ade}, P.~A.~R., {Aghanim}, N., {et~al.} 2016, \aap,
  594, A13, \dodoi{10.1051/0004-6361/201525830}

\bibitem[{Rodriguez \& Loeb(2018)}]{Rodriguez:2018rmd}
Rodriguez, C.~L., \& Loeb, A. 2018, Astrophys. J. Lett., 866, L5,
  \dodoi{10.3847/2041-8213/aae377}

\bibitem[{{Rodriguez} {et~al.}(2019{\natexlab{a}}){Rodriguez}, {Zevin},
  {Amaro-Seoane}, {Chatterjee}, {Kremer}, {Rasio}, \&
  {Ye}}]{2019PhRvD.100d3027R}
{Rodriguez}, C.~L., {Zevin}, M., {Amaro-Seoane}, P., {et~al.}
  2019{\natexlab{a}}, \prd, 100, 043027, \dodoi{10.1103/PhysRevD.100.043027}

\bibitem[{{Rodriguez} {et~al.}(2019{\natexlab{b}}){Rodriguez}, {Zevin},
  {Amaro-Seoane}, {Chatterjee}, {Kremer}, {Rasio}, \& {Ye}}]{Rodriguez2019}
---. 2019{\natexlab{b}}, \prd, 100, 043027, \dodoi{10.1103/PhysRevD.100.043027}

\bibitem[{{Romero-Shaw} {et~al.}(2020){Romero-Shaw}, {Kremer}, {Lasky},
  {Thrane}, \& {Samsing}}]{2020arXiv201114541R}
{Romero-Shaw}, I.~M., {Kremer}, K., {Lasky}, P.~D., {Thrane}, E., \& {Samsing},
  J. 2020, arXiv e-prints, arXiv:2011.14541.
\newblock \doarXiv{2011.14541}

\bibitem[{{Roulet} {et~al.}(2020){Roulet}, {Venumadhav}, {Zackay}, {Dai}, \&
  {Zaldarriaga}}]{2020arXiv200807014R}
{Roulet}, J., {Venumadhav}, T., {Zackay}, B., {Dai}, L., \& {Zaldarriaga}, M.
  2020, arXiv e-prints, arXiv:2008.07014.
\newblock \doarXiv{2008.07014}

\bibitem[{{Roulet} \& {Zaldarriaga}(2019)}]{2019MNRAS.484.4216R}
{Roulet}, J., \& {Zaldarriaga}, M. 2019, \mnras, 484, 4216,
  \dodoi{10.1093/mnras/stz226}

\bibitem[{{Safarzadeh} \& {Farr}(2019)}]{2019ApJ...883L..24S}
{Safarzadeh}, M., \& {Farr}, W.~M. 2019, \apjl, 883, L24,
  \dodoi{10.3847/2041-8213/ab40bd}

\bibitem[{{Salvatier} {et~al.}(2016){Salvatier}, {Wiecki{\^a}}, \&
  {Fonnesbeck}}]{2016ascl.soft10016S}
{Salvatier}, J., {Wiecki{\^a}}, T.~V., \& {Fonnesbeck}, C. 2016, {PyMC3: Python
  probabilistic programming framework}.
\newblock \doeprint{1610.016}

\bibitem[{{Samsing}(2018)}]{2018PhRvD..97j3014S}
{Samsing}, J. 2018, \prd, 97, 103014, \dodoi{10.1103/PhysRevD.97.103014}

\bibitem[{{Santoliquido} {et~al.}(2020){Santoliquido}, {Mapelli}, {Bouffanais},
  {Giacobbo}, {Di Carlo}, {Rastello}, {Artale}, \&
  {Ballone}}]{2020ApJ...898..152S}
{Santoliquido}, F., {Mapelli}, M., {Bouffanais}, Y., {et~al.} 2020, \apj, 898,
  152, \dodoi{10.3847/1538-4357/ab9b78}

\bibitem[{{Theano Development Team}(2016)}]{2016arXiv160502688short}
{Theano Development Team}. 2016, arXiv e-prints, abs/1605.02688.
\newblock \url{http://arxiv.org/abs/1605.02688}

\bibitem[{{Tiwari}(2020)}]{2020arXiv201208839T}
{Tiwari}, V. 2020, arXiv e-prints, arXiv:2012.08839.
\newblock \doarXiv{2012.08839}

\bibitem[{{Vink} {et~al.}(2020){Vink}, {Higgins}, {Sander}, \&
  {Sabhahit}}]{2020arXiv201011730V}
{Vink}, J.~S., {Higgins}, E.~R., {Sander}, A. A.~C., \& {Sabhahit}, G.~N. 2020,
  arXiv e-prints, arXiv:2010.11730.
\newblock \doarXiv{2010.11730}

\bibitem[{Virtanen {et~al.}(2020)Virtanen, Gommers, Oliphant, Haberland, Reddy,
  Cournapeau, Burovski, Peterson, Weckesser, Bright, {van der Walt}, Brett,
  Wilson, Millman, Mayorov, Nelson, Jones, Kern, Larson, Carey, Polat, Feng,
  Moore, {VanderPlas}, Laxalde, Perktold, Cimrman, Henriksen, Quintero, Harris,
  Archibald, Ribeiro, Pedregosa, {van Mulbregt}, \& {SciPy 1.0
  Contributors}}]{2020SciPy-NMeth}
Virtanen, P., Gommers, R., Oliphant, T.~E., {et~al.} 2020, Nature Methods, 17,
  261, \dodoi{10.1038/s41592-019-0686-2}

\bibitem[{{Vitale} {et~al.}(2019){Vitale}, {Farr}, {Ng}, \&
  {Rodriguez}}]{2019ApJ...886L...1V}
{Vitale}, S., {Farr}, W.~M., {Ng}, K. K.~Y., \& {Rodriguez}, C.~L. 2019, \apjl,
  886, L1, \dodoi{10.3847/2041-8213/ab50c0}

\bibitem[{Waskom \& the seaborn~development team(2020)}]{waskom2020seaborn}
Waskom, M., \& the seaborn~development team. 2020, mwaskom/seaborn, latest,
  Zenodo, \dodoi{10.5281/zenodo.592845}

\bibitem[{{Weatherford} {et~al.}(2021){Weatherford}, {Fragione}, {Kremer},
  {Chatterjee}, {Ye}, {Rodriguez}, \& {Rasio}}]{2021arXiv210102217W}
{Weatherford}, N.~C., {Fragione}, G., {Kremer}, K., {et~al.} 2021, arXiv
  e-prints, arXiv:2101.02217.
\newblock \doarXiv{2101.02217}

\bibitem[{{Yang} {et~al.}(2020){Yang}, {Bartos}, {Haiman}, {Kocsis},
  {M{\'a}rka}, \& {Tagawa}}]{2020ApJ...896..138Y}
{Yang}, Y., {Bartos}, I., {Haiman}, Z., {et~al.} 2020, \apj, 896, 138,
  \dodoi{10.3847/1538-4357/ab91b4}

\bibitem[{{Zevin} {et~al.}(2020){Zevin}, {Bavera}, {Berry}, {Kalogera},
  {Fragos}, {Marchant}, {Rodriguez}, {Antonini}, {Holz}, \&
  {Pankow}}]{2020arXiv201110057Z}
{Zevin}, M., {Bavera}, S.~S., {Berry}, C. P.~L., {et~al.} 2020, arXiv e-prints,
  arXiv:2011.10057.
\newblock \doarXiv{2011.10057}

\end{thebibliography}
\bibliographystyle{aasjournal}

\end{document}